\documentclass[review]{elsarticle} 

\usepackage{graphicx}
\usepackage{float}
\usepackage{tikz}
\usepackage{amsmath}
\usepackage{algorithm}
\usepackage{algpseudocode}
\usepackage{epstopdf}
\usepackage{tikz}
\usepackage{tcolorbox}
\usepackage{subfig}
\usepackage{balance}
\usepackage{color}
\usepackage{amssymb}
\usepackage{multirow}
\newcommand{\todo}[1]{{\color{black} #1}}

\newcommand{\revise}[1]{{\color{black} #1}}

\newcommand{\todoo}[1]{{\color{black} #1}}
\begin{document}

\title{Empirical Evaluation of the Heat-Diffusion Collection Protocol for Wireless Sensor Networks}

\author[usc]{Pradipta Ghosh\corref{cor1}}
\ead{pradiptg@usc.edu}
\author[focal]{He Ren}
\ead{rhilogin@gmail.com}
\author[usc]{Reza Banirazi}
\ead{banirazi@gmail.com}
\author[usc]{Bhaskar Krishnamachari}
\ead{bkrishna@usc.edu}
\author[usc]{Edmond Jonckheere}
\ead{jonckhee@usc.edu}

\cortext[cor1]{Corresponding author}
\address[usc]{Ming Hsieh Department of Electrical Engineering, University of Southern California, Los Angeles, California, USA}
\address[focal]{Electrical Engineering, Stanford University, Stanford, California, USA}

\begin{abstract}
Heat-Diffusion (HD) routing is our recently-developed queue-aware routing policy for multi-hop wireless networks inspired by Thermodynamics. In the prior theoretical studies, we have shown that HD routing guarantees throughput optimality, minimizes a quadratic routing cost, minimizes queue congestion on the network,  and provides a trade-off between routing cost and queueing delay that is Pareto-Optimal. While striking, these guarantees are based on idealized assumptions (including global synchronization, centralized control, and infinite buffers) and heretofore have only been evaluated through simplified numerical simulations. \revise{We present here the first practical decentralized version of HD algorithm, which we refer to as Heat-Diffusion Collection Protocol (HDCP), for wireless sensor networks.} We present a thorough evaluation of HDCP based on real testbed experiments, including a comparative analysis of its performance with respect to the state of the art Collection Tree Protocol (CTP) and Backpressure Collection Protocol (BCP) for wireless sensor networks. We find that HDCP has a significantly higher throughput region and greater resilience to interference compared to CTP. However, we also find that the best performance of HDCP is comparable to the best performance of BCP, due to the similarity in their neighbor rankings, which we verify through a Kendall's-Tau test. 
\end{abstract}


\maketitle

%

\section{Introduction}
\label{sec:intro}

Low power wireless sensor networks tend to be used for low-data rate applications. However, their scaling in terms of network size as well as operation under low duty cycles is often limited due to bandwidth constraints. Routing algorithms that can utilize the full bandwidth capacity of the network are therefore very important and continue to be a subject of research and development.  

\textbf{Throughput Optimality: } In network theory, the ability to fully utilize the available bandwidth in a network is tied to the notion of throughput optimality. \emph{An algorithm is said to be throughput optimal if it has the ability to maintain stable queues at any set of arrival rates that could possibly be stabilized by at least one algorithm.} The Back-Pressure (BP) routing algorithm~\cite{tassiulas1992stability} was the first queue-aware routing protocol to offer in theory a throughput optimality guarantee under general link state and traffic conditions.
It has been translated to practice in the form of the Backpressure Collection Protocol (BCP) ~\cite{moeller2010routing} for wireless sensor networks, which was shown to provide improved capacity and robustness to link dynamics compared to the state-of-the-art queue-unaware tree-based routing protocols \todo{(e.g., the well known Collection Tree Protocol, CTP~\cite{gnawali2009collection}).}

\textbf{What is Heat Diffusion (HD) algorithm?} The Heat Diffusion (HD) algorithm \cite{banirazi2014heat} is \todoo{our} recently proposed alternate queue-aware throughput optimal routing policy for wireless networks. It is derived from a combinatorial analog of the classical Heat Diffusion equation in Thermodynamics (where queue size is analogous to temperature, and packet flow to heat flow) that takes into account wireless interference constraints. Moreover, \todoo{in \cite{banirazi2014dirichlet} we have shown} that the underlying mathematical formalism is also essentially the same as current flows in resistive circuits. Therefore, the link penalties corresponding to resistances can be incorporated into HD routing in a way that allows for minimizing a specific form of average routing cost referred to as the Dirichlet routing cost. \emph{The Dirichlet routing cost is defined as the product of a link's cost and square of the respective link's flow rate.}
Moreover, the HD algorithm also minimizes the overall queue congestion of the network among the class of throughput optimal algorithms that make decision based on only current queue occupancies and link statistics. The HD routing algorithm guarantees to operate on the Pareto boundary if both routing costs and queue occupancies are considered in the objective function.
We detail the HD algorithm in Section~\ref{sec:HD}.

\textbf{Motivation of Our Work:} In theory, the HD routing algorithm goes beyond just throughput optimality guarantees to provide additional significant improvements in average queue sizes (delay) and average routing costs (such as expected transmission count, ETX\footnote{The expected transmission count (ETX) is a well-known link quality metric in the context of wireless communication which represents the average number of transmissions required for every successful packet transfer~\cite{de2005high}.}) compared to traditional Backpressure routing.
To date, this HD algorithm has remained a theoretical and idealized construct that requires a centralized implementation.
This centralized version requires a complete knowledge of a network and a NP-hard scheduling procedure at each time and assumes that buffer sizes are unbounded at all nodes. What has been missing in the literature is a practical implementation of the HD policy that is distributed and works with finite buffer lengths, and whose performance is studied comprehensively on a real wireless testbed. We seek to address this gap.

\textbf{Our Contribution: }
Our contribution in this paper is multi-fold. \textbf{First}, we present the first-ever decentralized version of the Heat-Diffusion algorithm and present a Contiki OS~\cite{dunkels2011contiki} based practical protocol implementation for data collection in wireless sensor networks: the Heat Diffusion  Collection Protocol (HDCP). This also include practically-motivated enhancements of the original Heat Diffusion algorithm, modifications to the link weight calculations, and a link switching scheme to diversify the link usage.
\textbf{Second}, we propose and evaluate a new method of dynamic ETX calculation suitable for any dynamic routing algorithm, including the previously proposed BCP~\cite{moeller2010routing} as well as HDCP. 
\textbf{Third}, we present and analyze the data collected from an extensive set of practical experiments conducted with HDCP utilizing forty five nodes on a real wireless sensor network testbed. Based on these data, we discuss the variation in the performance of HDCP under different parameters. 
\textbf{Fourth}, we compare HDCP with a Contiki-OS implementation of BCP~\cite{moeller2010routing} as well as CTP~\cite{gnawali2009collection}. We show that on the real testbed, HDCP offers significant improvements in performance over CTP in terms of throughput as well as resilience to external interference and node failures. We also show that the performance of HDCP is similar to BCP, and through evaluation of a Kendall's Tau similarity measure, show that this is due to similar rankings among the neighbors. 
\textbf{Finally}, we also verify that HDCP performs well with a low power communication stack (CX-MAC, a version of X-MAC\cite{buettner2006x} that is provided in Contiki, with 5$\%$ duty cycle).

\textbf{Paper Organization:}
The rest of the paper is organized as follows. In Section~\ref{sec:related}, we present a brief overview of the existing related works \todo{followed by a brief summary of the theoretical HD algorithm in Section~\ref{sec:back}.
We introduce the proposed Heat Diffusion Collection Protocol in Section~\ref{sec:hdcp}. }
Section~\ref{sec:implement} explains the practical implementation details of HDCP for the real experiment based comparative analysis of HDCP presented in Section~\ref{sec:real_exp}. The similar empirical performance between HDCP and BCP is analyzed and explained in Section~\ref{sec:similarity}. Section~\ref{sec:concl} concludes the paper.

\section{Related Works}
\label{sec:related}
Besides the original Backpressure routing algorithm, other throughput optimal policies \cite{dai2008asymptotic,shah2006optimal,naghshvar2012general} have also been proposed in the existing network theory literature. \todoo{The HD algorithm also provides the same throughput optimality guarantee in theory. However, what motivated us to implement HD were the striking additional expected performance capabilities (based on our theoretical results)--- that it also offers a Pareto-optimal trade-off between routing cost and queue congestion.}

There have also been several reductions of Backpressure routing to practice in the form of distributed  protocols, pragmatically implemented and empirically evaluated for different types of wireless networks~\cite{moeller2010routing,Martinez11,Alresaini12}. Most relevant to the present work is the Backpressure Collection Protocol (BCP) developed by Moeller \emph{et al.}~\cite{moeller2010routing}, the first ever implementation of dynamic queue-aware routing in wireless sensor networks. Our present work is informed by the BCP approach to implement Backpressure routing in a distributed manner, and we also directly compare the performance of the new HDCP protocol with BCP. 

Besides BCP, there are a number of other prior works on routing and collection protocols for wireless sensor networks, including the Collection Tree Protocol (CTP)~\cite{gnawali2009collection}, Glossy~\cite{ferrari2011efficient}, Dozer~\cite{burri2007dozer}, Low-power Wireless Bus~\cite{ferrari2012low}, ORW~\cite{landsiedel2012low} and Oppcast~\cite{mohammad2016oppcast}. We provide a side by side comparison of HDCP with the well-known CTP and BCP protocols. We believe this provides a meaningful comparison with a state of the art minimum cost quasi-static routing protocol as well as a state of the art queue and cost-aware dynamic routing protocol.

In recent years there has been a significant focus in developing networking protocols that are IP-friendly, such as RPL~\cite{RPL}. While the present paper does not focus on providing an IP-compliant version of HD, there is prior work on extending BCP to handle IP packets~\cite{BackIP}, and we believe that a similar approach could be adopted to enable IP operation for HDCP in the future. 

\todoo{ In our prior works, we have presented the idealized Heat Diffusion routing algorithm~\cite{banirazi2014heat,banirazi2014dirichlet}.} These are network theory papers that spell out a centralized algorithm, assume global synchronization, assume that at each time step a NP-hard Maximum Weight Independent Set problem can be solved, and that all queues are of unlimited size, and under these assumptions prove various properties of the HD algorithm. The only evaluations presented in these works are idealized MATLAB simulations. \todoo{This work is clearly inspired by and built up on our earlier works on HD routing but is the first to develop and implement it as a realistic distributed protocol (HDCP) and evaluate it on a real testbed. }

\section{The Heat Diffusion Routing: Theory and Concepts}
\label{sec:HD}
\label{sec:back}


The general idea behind dynamic queue-aware routing algorithms such as the Backpressure~\cite{tassiulas1992stability} algorithm and the Heat Diffusion~\cite{banirazi2014heat} algorithm is that they do not require any explicit path computations.
Instead, the next-hop for each packet depends on queue-differential weights that are functions of the local queue size information and link state information at each node.
It is a general assumption in (theoretical) queue-aware routing algorithms such as BP and HD that the networks operate in slotted time.
Furthermore, a wireless network is represented as a graph $(\mathcal{V}, \mathcal{E})$ with vertices $\mathcal{V}$ and edges $\mathcal{E}$.
However, the adjacent links or edges of a wireless network cannot be used simultaneously due to many constraints such as interference.
In that context, a \emph{maximal schedule} is defined as a set of links such that no two links interfere with each other and no other link can be added to that set without causing interference.
We will denote a maximal schedule as: $\pi=\{ \pi_{ij}| i\not = j \mbox{\ and \ } i,j\in \mathcal{V}\} \in \{0,1\}^{|{\mathcal{E}}|}$, where $\pi_{ij} = 1$ if the link $ij$ (or link $ji$) is included in the schedule. 
The set of all such maximal schedule is referred to as a \emph{scheduling set}, denoted as $\Pi$.

The Heat Diffusion (HD)~\cite{banirazi2014heat} routing has been derived from the combinatorial analogue of classical Heat-Diffusion equation in Thermodynamics.\
It uses the information about estimated link capacities, $\mu_{ij}(n)$, link cost factor, $\rho_{ij}(n)$, and queue backlogs, ${q}_i(n)$  $\forall i \in \mathcal{V}$, to make the routing decisions to put  $f_{ij}$ packets at each time slot $n$.
The optimization goal of the HD algorithm for a wireless network in theory can be described as follows~\cite{banirazi2014heat,banirazi2014dirichlet}:
\begin{equation}
\label{eqn:pareto}
\begin{split}
   \mbox{Minimize:}\ &(1-\beta)\overline{Q}+\beta \overline{R}\\
  \mbox{Subject to:}\ &\mbox{(1) Throughput optimality, and}\\
  &\mbox{(2) Network constraints}
\end{split}
\end{equation}
where $\overline{R}=\sum_{ij\in \mathcal{E}}\overline{\rho_{ij}(f_{ij})^2}$ is the Dirichlet average routing cost, $\overline{Q}=\sum_{i\in \mathcal{V}}\overline{q_{i}}$ is the average network queue size, and $\beta \in [0,1]$ is the control parameter to determine the trade-off between these two optimization goals. Note that $\beta$ is the only controllable parameter in the HD formulation. \emph{Throughput optimality} for a routing algorithm refers to its ability to maintain all queues stable for all sets of arrival rates for which it is possible by an omniscient router to maintain stable queues. Network constraints include constraints on link rates as well as interference constraints.
\todo{
Next, we give more concrete details on the HD routing algorithm along with a side by side comparison to the BP routing algorithm first proposed by Tassiulas and Ephremides~\cite{tassiulas1992stability} and extended by Neely \emph{et al.}~\cite{georgiadis2006resource,neely2010stochastic}. 
For clarity, we summarize the symbols used in Table~\ref{table:symbol}.
The HD routing algorithm as well as the BP routing algorithm has three steps that can be described as follows.
\begin{table}[H]
\centering
\caption{List of Symbols}
\label{table:symbol}
\resizebox{0.9\linewidth}{!}{
\begin{tabular}{|c|c|c|}
    \hline
    \multicolumn{3}{|c|}{Variables}\\
    \hline
    Symbol &  Value Set & Description \\
     \hline
    $q_i$ & $\mathbb{Z}_{\geq0}$ & The Queue Backlog at Node $i$ \\ 
    $q_{ij}$ &  $\mathbb{Z}$ & The Queue Differential Between Node $i$ and Node $j$\\ 
    $\mu_{ij}$ & $\mathbb{R}^+_0$ & Capacity of Link $ij$\\
    $\omega_{ij}$ & $\mathbb{R}$ & The Calculated Weight of Link $ij$\\ 
    $f_{ij}$ & $\mathbb{Z}_{\geq0}$ & The Number of Packets to be Sent for HD \\
     \hline
     \hline
    \multicolumn{3}{|c|}{Parameters}\\
    \hline
    Symbol &  Value Set & Description \\
     \hline
     $V$ & $[0,\infty)$ & Penalty Weighing Parameter in BP\\ 
     $\rho_{ij}$ & $\mathbb{R}$ & Penalty for link $ij$ in BP\\
     $\beta$ & $[0,1]$ & Pareto Optimal Trade-off Control Parameter for HD \\
     \hline
\end{tabular}}
\end{table}

\subsection{Link Weighing}

\textbf{HD: }
 Calculate the number of packets the link will transmit if it is activated at time-slot $n$.\
 In the original literature, this quantity is denotes by $\widehat{f_{ij}}(n)$, calculated as follows:
 \begin{equation}
 \label{eqn:hd_1}
\begin{split}
\widehat{f_{ij}}(n) &= \min \{ \phi_{ij}(n)q_{ij}(n)^+ , \mu_{ij}(n) \} \\
\phi_{ij}(n) &= (1-\beta) +\beta/\rho_{ij}(n)
\end{split}
\end{equation}
{where the Lagrange parameter $\beta$ is defined in Eqn~\eqref{eqn:pareto} }
Now, the link weights are calculated as follows:
\begin{equation}
w_{ij}(n) = 2\phi_{ij}(n)q_{ij}(n)\widehat{f_{ij}}(n)-\widehat{f_{ij}}(n)^2
\end{equation}

\textbf{BP: }
In the original BP routing, which only considers queue stabilities, link weights are calculated as follows:
\begin{equation}
w_{ij}(n)= \mu_{ij}(n) q_{ij}(n)
\end{equation}
To incorporate the routing cost into the BP,
the drift-plus-penalty approach~\cite{georgiadis2006resource,neely2010stochastic} was proposed, which we refer to as the V-parameter BP algorithm.
In this approach, a route usage cost is added as a negative penalty in the weight calculation as follows:
\begin{equation}
\label{v_param_cost}
w_{ij}(n)= \mu_{ij}(n) ( q_{ij}(n) - V .\rho_{ij}(n))
\end{equation}
where $V\in [0,\infty)$ determines the importance of the link penalty, and $\rho_{ij}(n)$ is the link penalty which depends on the link utility or cost function along with some penalty functions.

\subsection{Scheduling}

\textbf{HD \& BP: }
Find a scheduling vector $\pi \in \Pi$ such that:
\begin{equation}
\Gamma(n) = \arg\underset{\pi \in \Pi}{\max{}} \sum_{ij \in \mathcal{E}}\pi_{ij}w_{ij}(n)
\end{equation}
and the ties are broken randomly.


\subsection{Forwarding}

\textbf{HD: }
At this step, send $\widehat{f_{ij}}(n)$ number of packets over the link $ij$ if $\pi_{ij}(n)=1$ and $w_{ij}(n)>0$.
However, $\widehat{f_{ij}}(n)$ may be a fractional number.\
Therefore the actual number of packets transmitted is:  
\begin{equation}
f_{ij}(n)=\begin{cases} \lceil \widehat{f_{ij}}(n) \rceil & \mbox{if $\pi_{ij}(n)=1$} \\
0 & \mbox{otherwise}
\end{cases}
\end{equation}

\textbf{BP: }
Based on the scheduling vector from the previous step i.e., $\Gamma(n)$, if a link is active at time-slot $n$ i.e., $\pi_{ij}(n)=1$, and if the link weight $w_{ij}(n) > 0$, then transmit packets on that link at full capacity $\mu_{ij}(n)$.
Null packets are sent if a node does not have enough packets to send.

\emph{Note that, unlike BP routing, a node running HD routing sends $f_{ij}(n)$ number of packets rather than transmitting at the full capacity, $\mu_{ij}(n)$.}
}
In Table~\ref{tab:contrast}, we summarize the side by side comparison of the BP and the HD algorithms. For a more detailed comparative analysis of the theoretical BP and HD algorithms, interested readers are referred to the original HD paper~\cite{banirazi2014heat}.

\newsavebox\unistrain
\begin{lrbox}{\unistrain}
  \begin{minipage}{0.7\textwidth}
    \begin{equation*} f_{ij}(n)=\begin{cases} \lceil \widehat{f_{ij}}(n) \rceil & \mbox{if $\pi_{ij}(n)=1$} \\
        0 & \mbox{otherwise} \end{cases} \end{equation*}
  \end{minipage}
\end{lrbox}

\begin{table}[!h]
    \centering
    \caption{Contrasting HD policy with V-parameter BP policy (adapted from~\cite{banirazi2014heat})}
    \begin{tabular}{|c|ccc|}
    \hline
        \multirow{4}*{Weighing} & \multirow{2}*{$\widehat{f_{ij}}(n)$} & ~~BP~ & $\min\bigl\{ \,\mu_{ij}(n),\: q_i(n) \bigr\}$ \\  \cline{3-4}
        && ~~HD~          & $\min\bigl\{ \bigl( 1\! - \!\beta + \beta/\rho_{ij}(n) \bigr) q_{ij}(n)^+\! ,\: \mu_{ij}(n) \bigr\}$ \\  \cline{2-4} 
        & \multirow{2}*{~~~$w_{ij}(n)$} & ~~BP~ & $\mu_{ij}(n) \bigl(q_{ij}(n)-V\rho_{ij}(n)\,\mu_{ij}(n) \bigr){^+}$ \\ \cline{3-4}
        && ~~HD~          & ~~$2\bigl( 1\! - \!\beta + \beta/\rho_{ij}(n) \bigr)
                         q_{ij}(n)\widehat{f_{ij}}(n) - \widehat{f_{ij}}(n)^2$ \\ \hline
                         
        {Scheduling} & \multicolumn{3}{c|}{$\Gamma(n) = \arg\underset{\pi \in \Pi}{\max{}} \sum_{ij \in \mathcal{E}}\pi_{ij}w_{ij}(n)$} \\ \hline
        Forwarding & \multicolumn{3}{c|}{\usebox{\unistrain}} \\ 
    \hline
    \end{tabular}
    \label{tab:contrast}
\end{table}

\section{The Heat Diffusion Collection Protocol : From Theory to Reality}
\label{sec:hdcp}
\todo{
In this section, we detail the Heat Diffusion Collection Protocol (HDCP) and the modifications made to the theoretical HD protocol for practical implementation.

\subsection{Predecessors:}
Before detailing the HDCP, we present a brief overview of two of the well known data collection routing protocols for a side by side comparison: Backpressure Collection Protocol (BCP) and Collection Tree Protocol (CTP). Moreover, these two protocols gave us insights on molding the promising theoretical HD algorithm into a real implementations.

\subsubsection{The Collection Tree Protocol}  
The Collection Tree Protocol (CTP) \cite{gnawali2009collection} is a tree based, best-effort, anycast data collection protocol that was first introduced in \cite{fonseca2006collection}. 
There have been many practical implementations of CTP among which CTP Noe, presented in \cite{gnawali2009collection}, is the most popular one.
The main idea of CTP is to maintain minimum cost trees to a set of nodes that advertise themselves as the data sink/ tree roots. 
The distance/cost used in this context is in terms of the well-known metric called ETX. 
Each node calculates the shortest distance to a sink/root in terms of ETX and uses the respective next hop neighbor to send the data.
In CTP, there exist two types of packets: data packets and routing packets.
While the data packets are used for actual data transmission, the routing packets are used solely to setup/update the tree.
For setting up the tree, CTP uses a variant of the Trickle algorithm~\cite{levis2004trickle}.
CTP uses an adaptive beaconing technique to identify the neighbors, to calculate the shortest path, and to adapt after node failures or link quality changes.
To avoid routing loops, CTP uses a datapath validation technique.
In this technique, if a node receives a packet from a node with lower or equal distance/cost (in terms of ETX) to a root, it triggers a router repair phase and retry after a timeout.
\todo{In contrast, neither BCP nor HDCP relies on a predetermined routing path.}
%

\subsubsection{The Backpressure Collection Protocol}


The Backpressure Collection Protocol (BCP)~\cite{moeller2010routing} is a distributed dynamic routing protocol which practically implements the idealized V-parameter BP algorithm without the need for a global max weight scheduling.\
In this protocol, the link penalty, $\rho_{ij}(n)$, in \eqref{v_param_cost} is replaced by $\overline{ETX_{ij}}(n)$, which is the $ETX$ estimate for link $ij$ at time-slot $n$.
Therefore, the modified weighing function is as follows:
\begin{equation}
\label{v_param_cost1}
w_{ij}(n)= \mu_{ij}(n) ( q_{ij}(n) - V .\overline{ETX_{ij}}(n))
\end{equation}
In this distributed protocol, each node calculates the weight for each of its outgoing links locally and chooses the neighbor with the maximum positive weight, if any, to forward the next packet.\
It is shown in \cite{moeller2010routing} that a last-in-first-out (LIFO) queue implementation of BCP is better than first-in-first-out (FIFO) in terms of delay performance.\
Also, floating or virtual queues are implemented to deal with the problem of limited buffer size.
}
\subsection{The Heat Diffusion Collection Protocol: }

The original HD algorithm is a centralized protocol where at each time slot the optimum non-interfering schedule must be computed. In our distributed implementation of the Heat Diffusion Collection Protocol (HDCP), every node decides the next hop locally and greedily based on the weight calculations.
\emph{Moreover, following the logic of BCP the penalty/cost factor $\rho_{ij}(n)$ in \eqref{eqn:hd_1} is replaced by $\overline{ETX_{ij}}(n)$ which is the estimated $ETX$ of the link $ij$ at time-slot $n$.}
Thus, the modified equations to calculate the link weights are as follows:
\begin{equation}
 \label{eqn:hd_2}
\begin{split}
\widehat{f_{ij}}(n) &= \min \{ \phi_{ij}(n)q_{ij}(n)^+ , \mu_{ij}(n) \} \\
\phi_{ij}(n) &= (1-\beta) +\beta/\overline{ETX_{ij}}(n)
\end{split}
\end{equation}
\begin{equation}
\label{eq:hd}
\begin{split}
w_{ij}(n) = 2\{(1-\beta) +\beta/\overline{ETX_{ij}}(n)\}q_{ij}(n)\widehat{f_{ij}}(n)-\widehat{f_{ij}}(n)^2
\end{split}
\end{equation}
Now, each node calculates the weight for each of its outgoing links and chooses the link with the maximum positive weight. 
Note that, most of the variables in the weight calculation can be estimated or calculated during the operation if provided with a value of $\beta$.
We explain the choice of $\beta$ in the next section, followed by detailed descriptions of other components of the HDCP such as distributed weight calculation, link ETX estimation, and our proposed link switching method to improve the link qualities. 
\revise{
Moreover, unlike the centralized algorithm's NP-hard maximum weight independent set time scheduling to avoid interference, our distributed protocol handles interference by adaptive retransmissions and CSMA based MAC access.
}

\subsubsection{The $\beta$ Parameter} 
\label{sub:qpara}
In theory, for different choices of $\beta$, we should get different performance for HDCP  as the optimization goal changes for different values of $\beta$ \todo{(Note that this parameter is not part of the CTP and BCP formulations)}.
If we choose $\beta=0$, Eqn.~\eqref{eq:hd} will be simplified to:
\begin{equation}
\label{eq:hd1}
\begin{split}
\widehat{f_{ij}}(n) &= \min \{q_{ij}(n)^+ , 1 \}\\
w_{ij}(n) &= 2q_{ij}(n)\widehat{f_{ij}}(n)-\widehat{f_{ij}}(n)^2
\end{split}
\end{equation}
Now, for $q_{ij}(n)>0$, $\widehat{f_{ij}}(n)=1$ as the queue differential can take only integer values.
Therefore, Eqn~\eqref{eq:hd1} can be rewritten as follows:
\begin{equation}
\label{eq:hd1_simple}
 w_{ij}(n) =
  \begin{cases}
   {2q_{ij}(n) -1} & \text{if } q_{ij}(n) > 0 \\
   0       & \text{Otherwise } 
  \end{cases}
\end{equation}
Therefore, the optimization goal becomes similar to the goal in the original ``pure" BP routing (by Tassiulas and Ephremides~\cite{tassiulas1992stability}) and it does not include the minimization of ETX. 
Also, as the link weights solely depend on the queue differentials, the delay performance should be better \emph{provided that the links are all very good\footnote{A link is very good/perfect if the ETX of the link is 1 which implies that every packet transmission is successful on that link.}.}
On the contrary, since this protocol does not try to minimize the ETX, it can choose a bad link\footnote{We consider a link to be very bad if the ETX is $\geq 5$ i.e., one successful transmission per every five transmissions.} if the queue gradient on that link is the largest. 
As a consequence, the average number of retransmissions faced by a packet also increases which is directly translated to larger end to end delay. 
Therefore, if the overall path costs in terms of ETX is the dominant factor in the end to end delay calculation,  HDCP with $\beta=0$ may perform poorly in practice.

On the other hand, if $\beta=1$, Eqn.~\eqref{eq:hd} will be as follows:
\begin{equation}
\label{eq:hd2}
\begin{split}
\widehat{f_{ij}}(n) &= \min \{\frac{q_{ij}(n)^+}{\overline{ETX_{ij}}(n)} , 1 \}\\
w_{ij}(n) &= 2\frac{q_{ij}(n)}{\overline{ETX_{ij}}(n)}\widehat{f_{ij}}(n)-\widehat{f_{ij}}(n)^2
\end{split}
\end{equation}
Similar to the previous case, we can simplify~\eqref{eq:hd2} as follows:
\begin{equation}
\label{eq:hd2_simple}
 w_{ij}(n) =
  \begin{cases}
   2\left(\frac{q_{ij}(n)}{\overline{ETX_{ij}}(n)}\right)-1 & \text{if} \frac{q_{ij}(n)}{\overline{ETX_{ij}}(n)}\geq 1 \\
   \left( \frac{q_{ij}(n)}{\overline{ETX_{ij}}(n)}\right)^2 & \text{if } 0 < \frac{q_{ij}(n)}{\overline{ETX_{ij}}(n)} < 1 \\
   0 & \text{Otherwise }
  \end{cases}
\end{equation}
In this case the optimization goal is mainly the reduction of overall path costs in terms of ETX.
Thus the overall ETX for a path should be improved for this case.
However, this might result in a slight increase in hop counts if the links are very lossy.
The first and last cases in Eqn.~\eqref{eq:hd2_simple} correctly fulfill our routing requirement.
But the second case causes inefficiency in the real testbed experiments.
In such cases, even if the ETX cost is very high for a link and the queue differential is as low as 1, a node will try to send the packet to that link according to the original HD rule. 
Moreover, in practical experiments, the probability of falling under such a situation is very high.
Thus, it will negatively affect the overall performance of the HDCP and needs to be avoided.
Furthermore, provided that we have avoided any such situations and have a good link with $ETX=1$, a queue differential of $1$ will still result in a positive weight thereby causing the protocol to forward the packet. This results in a absence of a steady state queue gradient on such links. This can potentially increase the number of hops traversed by the packets and also deteriorates the goodput. 
In order to avoid both of these situations, we replace the $\rho_{ij}(n)$ in Eqn~\eqref{eqn:hd_1} by $\mathbf{V} \times \overline{ETX_{ij}}(n)$ which modifies Eqn.~\eqref{eq:hd} as follows:
\begin{equation}
\label{eq:hd2_mod1}
\begin{split}
w_{ij}(n) &= 2\{(1-\beta) +\frac{\beta }{\mathbf{V} \times \overline{ETX_{ij}}(n)}\}q_{ij}(n){f_{ij}}(n)-f_{ij}(n)^2\\
\end{split}
\end{equation}
where $f_{ij}(n)=\lceil \widehat{f_{ij}}(n) \rceil $.

By setting $\mathbf{V}\geq 2$, we make it certain that there exists a steady state queue gradient towards the sink.
Therefore, a node will consider a link only if ${q_{ij}(n)}$ is greater than $\overline{ETX_{ij}}(n)$.
Thus, for a link with very high ETX, the queue differential has to be higher in order to consider that link.
Furthermore, this strategy also satisfies the Backpressure criterion as for $q_{ij}(n)<0 \implies w_{ij}(n)<0$.
In Section~\ref{sec:unmod}, we present a practical experiment based analysis of the performance improvement as a result of this change in weight calculation.


\subsubsection{Updating Weights}
\label{sec:weight}
\todo{In order to calculate the weights in distributed manner, each node requires updated information about the queue sizes of its neighboring nodes without affecting the performance of the routing task.}
In our distributed implementation of HDCP, we employ two techniques to do that. \textbf{First,} during a long period of inactivity, each node periodically broadcasts a beacon with its current queue status similar to common wireless access points. If a neighboring node receives this broadcast, it will update its locally stored queue differential information. 
\textbf{Second,} when a node sends a data packet, it includes its current queue state in that packet's header. Due to the nature of wireless links, every packet is received by all the neighboring nodes {(we assume that no advanced MAC protocol is employed that schedules nodes to communicate in pairs at different times).} Once a data packet is received by a node, it sniffs the header of the packet to extract the queue information and updates the local queue information database, even if the respective node is not the destination of the packet. 

\todo{Note that, BCP follows similar technique for updating link weights. In CTP, there is no concept of queue differential based link weights. Rather, CTP uses beaconing based ETX information to predetermine the routing paths.}


\subsubsection{Queue Implementation}

\todo{\emph{Similar to BCP,} the practical implementation of the HDCP can have a FIFO queue or a LIFO queue implementation.} Based on the observation in~\cite{moeller2010routing} that LIFO queue implementation has a significantly better performance in terms of end-to-end delay (which we also observed empirically), we present only the LIFO queue implementation of the HDCP protocol in this paper. We also adopt the virtual ``floating" queue approach proposed in~\cite{moeller2010routing} to prevent packet buffer overflows due to the steady state queue gradient.

\subsubsection{Link Metric Estimation}
 One of our contributions in this paper is to propose a new method of ETX calculation for implementations of dynamic routing. \todo{\emph{Initially, we opted to follow the ETX calculation technique from original BCP paper~\cite{moeller2010routing}. In that implementation the estimation of $\overline{ETX_{ij}}$ for link $ij$ is performed in an online manner where the metric is updated by taking exponential weighted moving average of the number of retransmission attempts of the most recently transmitted packet.}}
This is a very effective way of ETX estimation for routing protocols that does not switch next hop during retransmission i.e., use the same link $ij$ for all the retransmission attempts.
However, for Backpressure-based dynamic routing protocols, the next hop calculation is performed before each retransmission, for path diversity. In such cases, we have to be very careful in calculating moving average since the same link may not be used for all the retransmissions. \todo{Therefore, an attempt to update the ETX for the most recently used link with the total number of retransmission attempts might lead to an erroneous ETX estimation.} To avoid this flaw, we can keep track of all the links used as well as the number of tries on that link and update either only the last used link or all the links after a successful packet transmission or a packet drop. 

As an alternative, we propose a 2-state discrete time Markov Chain based ETX estimation. In this method, we assume that each link can be either good (`1') or bad (`0') at certain point of time. With each state, we associate two transition probabilities: good to good ($p_{11}$), good to bad ($p_{10}$), bad to good ($p_{01}$) and bad to bad ($p_{00}$). 
Now the $\overline{ETX_{ij}}$ can be calculated as $\frac{1}{p_{01}}$ when the last state observed was a 0, and as $\frac{1}{p_{11}}$ when the last state was 1.


We maintain four counters associated with each routing table entry to keep track of different state transitions, denoted as $c_{00}$, $c_{01}$, $c_{10}$ and $c_{11}$. We also add a Boolean variable to keep track of the last state of the link, i.e., if the value is \textbf{true}, the last known state was good. We initialize the $c_{01}$ and $c_{11}$ to be $1$ and the others to be $0$. Now, every time a packet is transmitted (or retransmitted), the algorithm waits for a certain period of time to receive the acknowledgement (ACK). If received, the state of the link is set to be good (`1') otherwise it is set to be bad (`0'), and then based on the last state, the respective counter is increased by one. The counters may be reset after reaching a maximum value, to keep the ETX estimates fresh. In our experiments, we have not done this as it did not appear to affect the performance.




Now, based on the counters, the ETX is calculated (it can be shown that this corresponds to a maximum likelihood estimate of the underlying Markov Chain parameters) as follows:
\begin{equation}
\label{eq:etx2}
\overline{ETX_{ij}} =
\begin{cases}
   \frac{c_{00}+c_{01}}{c_{01}} & \text{if } \text{last State}=0 \\
   \frac{c_{10}+c_{11}}{c_{11}} & \text{if } \text{last State}=1
\end{cases}
\end{equation}

All the results presented in this paper are based on this new, more justifiable method of ETX calculation, which we apply to both BCP and HDCP for a fair comparison.

\subsubsection{Link Switching}
\label{sec:link_sw}
In this paper, we propose an enhancement of HDCP by introducing link switching. The main concept of link switching is to maintain a ordered set of best (in terms of the weights) K neighbors (K can be any positive integer) at each node. When a packet is sent, it is first sent to the first neighbor on this list. If transmission fails, the retransmission attempt is made immediately to the next neighbor in the list and so on. If the list is exhausted during retransmissions, the process restarts again from the first neighbor in the list. A node should fulfill some selection criterion to be included in the list such as the link weight should be within some threshold of the best link. In our experiments, we set a threshold on the ETX and weight i.e., if a positively weighted link's ETX is no worse than the best link's ETX + 1, we add that link to the list. 
In section \ref{sec:unmod}, we present a practical experiment based analysis of the performance improvement as a result of this change. 
However, we introduce this switching in HDCP only because we empirically found that it does not help to improve the performance of BCP.
\todo{Note that the concept of link switching is not part of the existing BCP and CTP implementations.}


\section{Implementation Details}
\label{sec:implement}
\todo{Similar to any data collection routing protocol, a number of common routing parameters need to be set properly in the real implementation of HDCP (in our case in the Contiki OS implementation) such as maximum queue size and maximum number of retransmissions.} In this section, we discuss the choices of such parameters and the reason behind them in details. First of all, we set the value of $\mu_{ij}(n)$ in Eqn.~\eqref{eqn:hd_1} to be 1 as a node cannot send more than one packet simultaneously. 


\subsection{Retransmission}
Retransmission is very crucial for the performance of any wireless network.\
For effective retransmission, the parameters such as retransmission timeout and maximum number of retransmissions have to be properly chosen.\
Retransmission is also directly related to the acknowledgement mechanism and the choice of ARQ.
Since the choice of ARQ affects the HDCP, the BCP and the CTP algorithms equally, we have implemented a simple Stop and Wait ARQ mechanism where a node can send only one packet at a time and wait for its acknowledgment before moving to the next packet. 
If the acknowledgement is not received within a certain time, commonly referred as retransmission timeout, the node retransmits the same packet.
Now, the value of this retransmission timeout directly affects the goodput of the system and needs to be properly chosen. \todoo{Note that, the ARQ mechanism is employed on top of the existing hardware level acknowledgement mechanism that tries a maximum of 3 times to properly transfer the packet to the next hop in case of unicast transmissions (e.g., software acknowledgements). We do not remove the hardware level acknowledgement (One key feature of the CTP algorithm) for a fair comparison as well as to avoid the unreliability issues in pure software acknowledgements.} 

In our experiments, the transmission and propagation time for a packet are in the order of tens of milliseconds. It would then perhaps be expected that the best setting for the retransmission should be on the order of around 10ms or so. Nevertheless, we empirically found that it is best to set the timeout for retransmitting a lost packet to be chosen randomly between 10 to 200 ms. We believe that this large range is needed because of the link coherence time in our testbed which is located in a busy office building environment. For instance, in~\cite{farahani2011zigbee}, it is indicated that the coherence time for IEEE 802.15.4 radios can be about 175ms. Retransmitting a lost packet quicker than the coherence time runs a higher risk of seeing another packet loss. Furthermore, we use CSMA/CA as the link access protocol, which also introduces some delay. 


The maximum number of retransmission attempts is set to 5 based on the original BCP code, which we empirically observed to perform well on our testbed.
After five retransmission attempts, if a packet is not acknowledged, the node will drop it and move to the next packet. 

\subsection{Retry}
Whenever a node generates or receives a packet, it tries to send it immediately (after about $4-5$ ms) if no other packet is being transmitted or waiting in the queue.\
However, when the node wants to transmit the packet, there might not be any suitable neighbor (in terms of having a positive weight) to forward the packet.\
In that case, the node needs to decide how much time should it wait before retrying. 
We refer to this wait time as the Retry time.
\todoo{One viable option is to constantly keep trying which is not efficient in terms of energy consumption due to radio wake times.}\
Also once this situation happens, it might take a while to have a good neighbor.
In this work, we set the retry time to be chosen randomly between 50ms to 100ms. The intuition behind choosing this value is again the transmission time for a packet being in the order of tens of milliseconds. 
Based on our experiments, we have also observed that the typical packet transfer time (the time duration between the transmission and reception of a packet) is $\approx 10ms$. Therefore, by choosing a value between 50ms and 100ms, we give the neighboring nodes enough time to potentially transfer several packets which is likely to be enough to create a positive weight.


\subsection{Queue Buffer}
In practical low power low memory devices, the possible queue buffer allocations are severely restricted. We fixed the maximum queue size to be 25 as this is the highest possible number of queue buffer that our device can accommodate alongside other required memories. Along with this buffer, there also exists a small memory allocated to store only the recent packet for the retransmission purpose.


\subsection{Beacon Timer}
Beaconing is a very important part of the practical implementation of both HDCP and BCP.\
When a node has nothing to send for long time, beacons are sent periodically, so that the neighboring nodes can keep their Backpressure database updated.
Also, beaconing is mandatory for a sink node since it has nothing to send.
Therefore we implement two different beaconing rates in our system.\
The first type of beaconing is for source nodes and the period for that is around 5 seconds.
The second type of beacons, which we refer to as the fast beacons, are used by the sink nodes and the period for that beacon is around 2 seconds. These values are chosen based on the original BCP code.

\subsection{Inbound Packet Filtering}
Inbound packet filtering is very important to improve the performance of both HDCP and BCP in the presence of retransmissions.
If no filtering is used, a node might receive multiple copies of the same packet due to retransmissions. Therefore the node might have multiple copies of the same packet stored in the buffer simultaneously, which is not efficient.
To avoid this kind of situations, we implement a inbound packet filter to drop any duplicate packets after sending proper acknowledgements.
In our implementation, each node maintains a history (packet source information and the origin sequence number) of 25 most recent packets received by the node. We choose this number to match the queue buffer size. Every time a node receives a packet, it checks the history, performs necessary action such as packet drop or store, and updates the history.

Further, to prevent packet looping, we implement a TTL counter which decrements at each hop. In our experiments, sources set the initial TTL for each packet conservatively to 10 (the maximum hop distance from any node to the sink in our testbed is only 3).  



\todoo{ 
\subsection{End to End Delay Calculations}
\label{sec:delay_cal}
For calculating end to end delay of each packet, we maintain a separate field called \textbf{$HDCP_{Delay}$} in the HDCP header, initialized with a value of $0$. At the source node, the packet is timestamped at the generation  (Say, $A_{source}$) and just before departure (Say, $D_{source}$), and the value $HDCP_{Delay}$ field is set to be $(D_{source}-A_{source})$. Similarly, we time-stamp the packet at each intermediate node, $I_k$: upon arrival ($A_{I_k}$) and just before departure ($D_{I_k}$); and add the time difference with the value of $HDCP_{Delay}$, i.e., $HDCP_{Delay}=HDCP_{Delay}+(D_{I_k}-A_{I_k})$. Thus, the value of the field $HDCP_{Delay}$ upon arrival on the sink denotes the end to end delay suffered by that packet. For illustration, assume that the travel path of a packet is $source \rightarrow I_1 \rightarrow \cdots \rightarrow I_M \rightarrow sink$, where $A_{I_1}, \cdots, A_{I_M}$ are the arrival times of the packet at the intermediate nodes, and $D_{I_1}, \cdots , D_{I_M}$ are the respective departure times. Then the end to end delay is: $ \sum_{i=1}^{i=M} |D_{I_i} - A_{I_i}| + |D_{Source}-  A_{source}|$. Note that, we do not add the propagation delays as the value of propagation delays are negligible compared to the queuing delay (which we measure) in our testbed setup.

}

\subsection{Experimental Setup}
\label{sec:experiment}

\begin{figure}[h]
    \centering
      \includegraphics[width=0.6\linewidth]{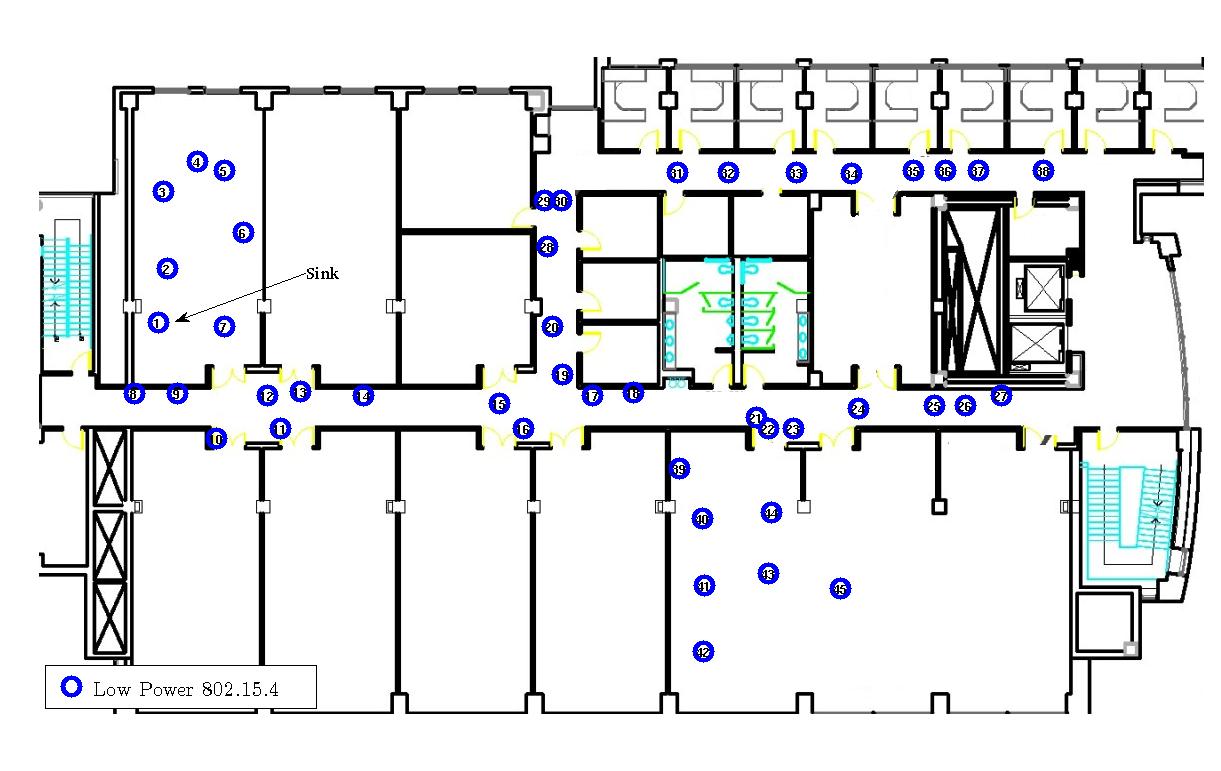}
     \caption{Real Experiment Testbed Topology}
    \label{fig:topology}
\end{figure}

To analyze the performance of HDCP in a real sensor network and compare it with BCP and CTP, we have implemented the HDCP and the BCP algorithms on Contiki OS and used the CTP implementation available with the Contiki OS. We perform a set of evaluation experiments on an indoor wireless network testbed called Tutornet \cite{tutornet} with forty five IEEE 802.15.4-based Tmote-sky nodes distributed over a floor with roughly $80,000$ sq.ft of area.
This testbed is also available for administered public use for approved research purposes including benchmarking protocols.
The network topology is presented in Figure~\ref{fig:topology} where the marked node is the sink and the rest of the nodes are the source nodes and the furthest node is three hops away from the sink.\
We use the channel number 26 with Tmote sky power level 31 for this purpose. 
The number of neighbors to each node varies from $19$ to $35$ with an average of $29$. 
Nonetheless, typically only about 7-8 of the neighbors are connected via good links ($ETX\approx 1$).
Thus the topology is very diverse with a considerable number of different paths between any two nodes in the network.
On the negative side, a considerable amount of interference exists among the nodes, which limits the bandwidth. 
The data packets in our experiments are all 26 Bytes in size.

All the experiments are performed on weekdays during daytime with lots of moving people and physical objects around. 
Each experiment is performed for 35 min: the network settles down during the first 5 min and the data is collected during the next 30 min. Each experiment is repeated at least 10 times to improve the confidence levels.
Note that, we discuss the experimental setup for low power stack and for node failures in Sections~\ref{sec:lowpower} and \ref{sec:failure}, respectively.

We evaluate HDCP's performance in terms of different values of $\beta$ and different packet generation rates.
We select the value of $V$ to be 2 for both BCP and HDCP which has been empirically determined to be an efficient operating point for BCP in the original BCP paper (which we could also verify in our own experiments).\

\section{Real Testbed Experiment Results and Analysis}
\label{sec:real_exp}

In this section, we evaluate the HDCP protocol under different configurations ($\beta$ values) and also compare them with LIFO BCP with virtual queue implementation and CTP.

\subsection{Variation of the $\beta$ Parameter}
\label{sec:beta_var}
We perform a set of practical experiments on the testbed with different values of $\beta$ and packet generation rates of 1 packet per 4 seconds per source (i.e., 0.25 PPS) as well as 1 packet per 2 seconds (0.5 PPS). The difference in performance is not prominent between the two source rates. Thus, we present the results only for the higher rate of 0.5 PPS in this section \todoo{to keep the length of the manuscript reasonable as well as to avoid presenting redundant information.}

The goodput of each source node is defined to be the number of packets received by the sink from it over a one second interval. For visual clarity, all the plots presented in this section are sorted in terms of the goodputs of the individual nodes for the experiment with $\beta=0$. The end to end delay calculation for each packet is performed by adding up all the queuing and processing delays in all intermediate nodes, \todoo{as discussed in Section~\ref{sec:delay_cal}}.
This ignores the propagation times which in any case are negligible compared to the processing delays.
\begin{figure}[!ht] 
 \centering
 \subfloat[]{\label{fig:beta:1}\includegraphics[height=0.4\linewidth,width=0.5\linewidth]{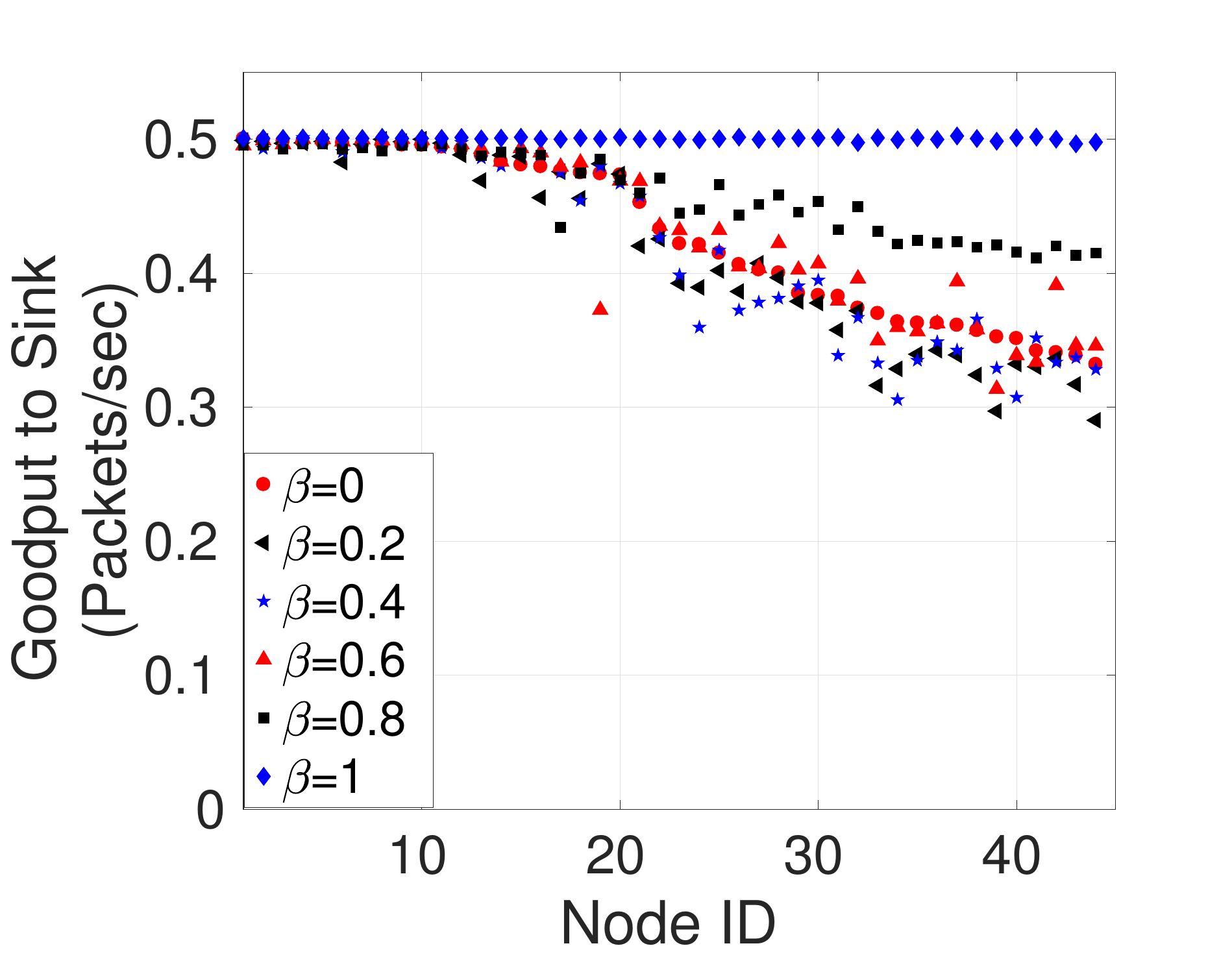}} 
 \subfloat[]{\label{fig:beta_cdf}\includegraphics[height=0.4\linewidth,width=0.5\linewidth]{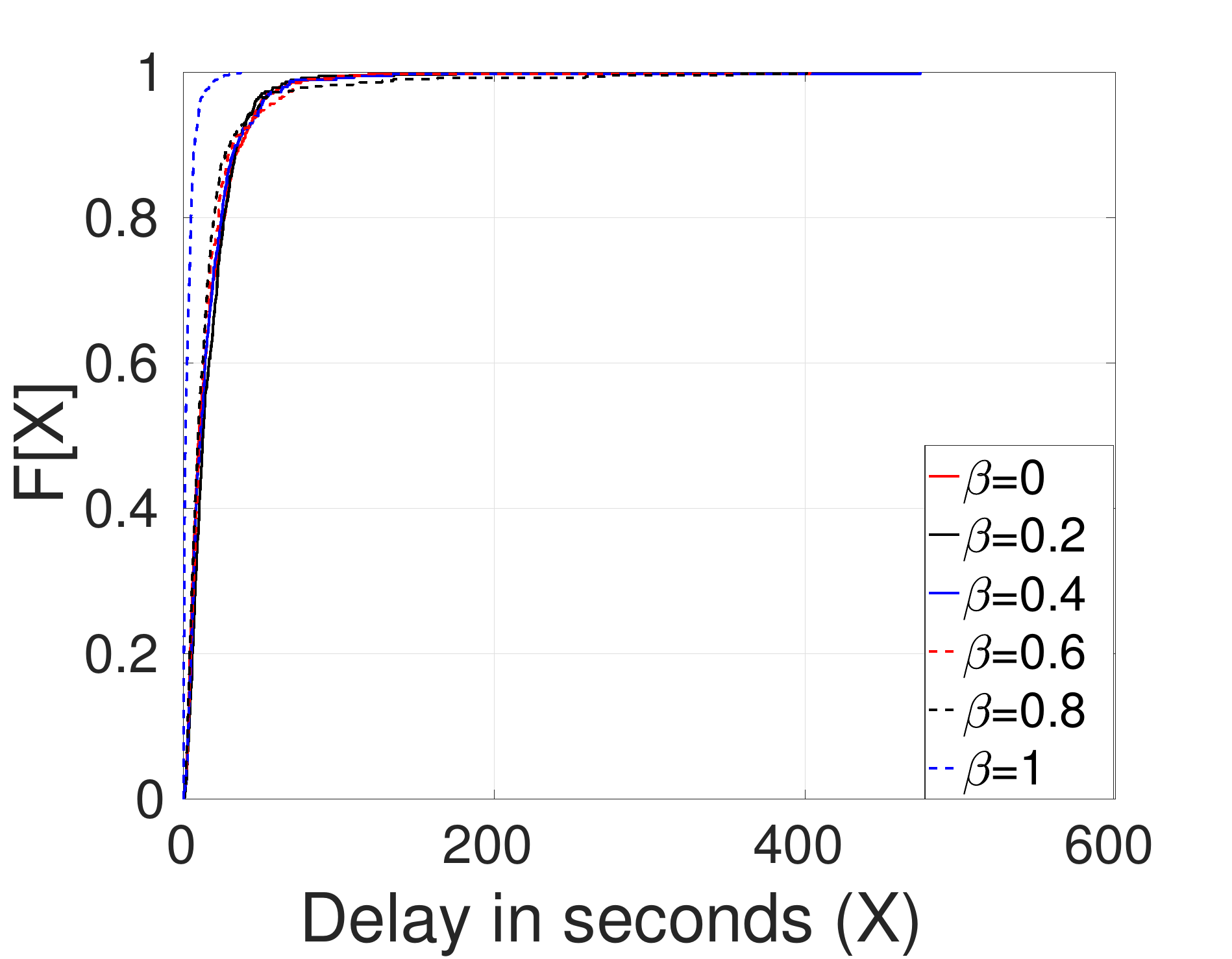}} \,
 \subfloat[]{\label{fig:beta:2}\includegraphics[height=0.5\linewidth,width=0.5\linewidth]{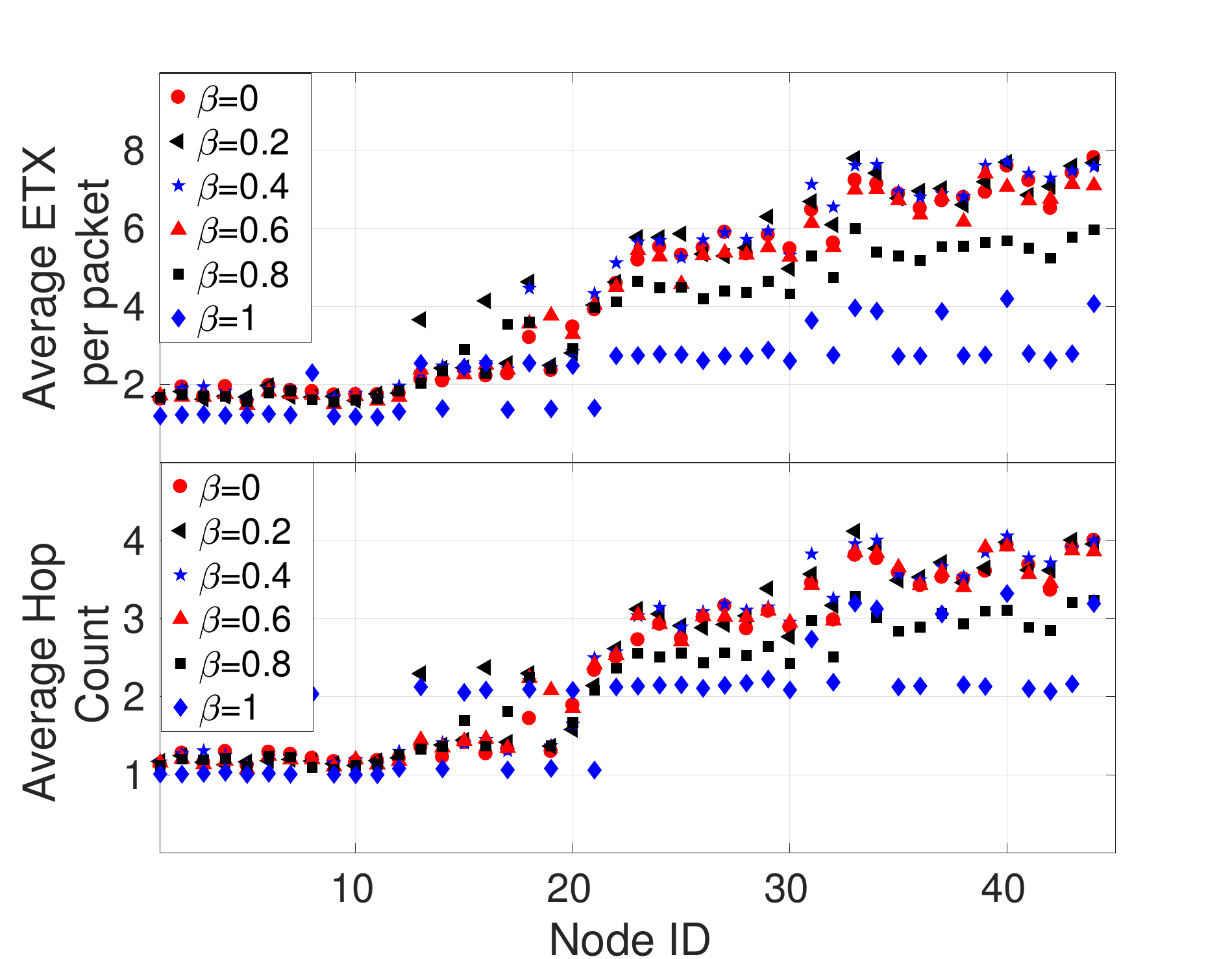}}
 \subfloat[]{\label{fig:beta1} \includegraphics[height=0.5\linewidth,width=0.5\linewidth]{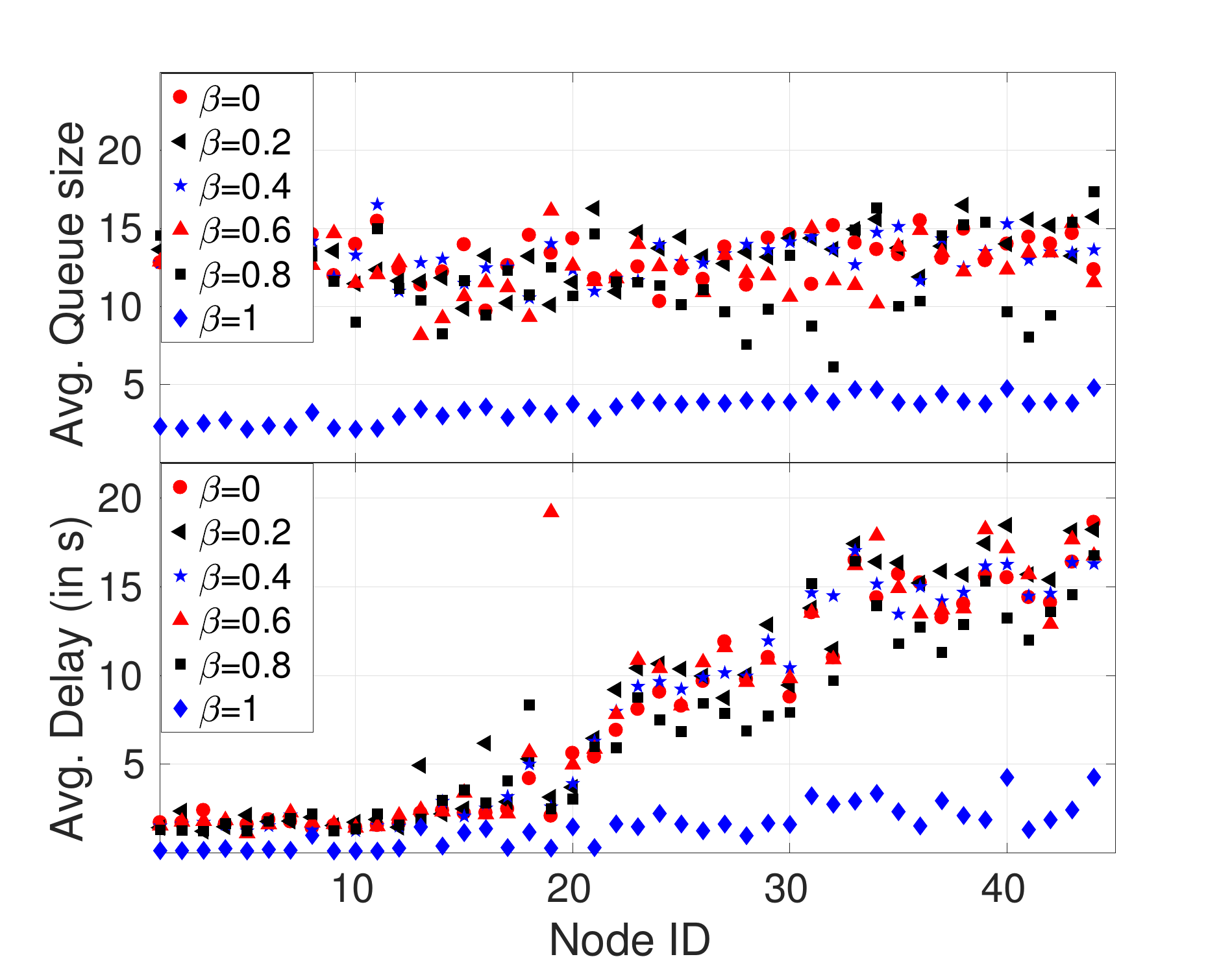}}
 \caption{Performance Plots of HDCP Implementation for 0.5 PPS with Different Values of $\beta$: (a) Average Goodput to Sink (b) End-to-End Delay CDF Plot for Mote 38 (c) Average ETX per Packet (Top) and Average Hop Count (Bottom) (d) Average End-to-End Delay (Bottom) and Average Queue Size for Each Node (Top)}
 \label{fig:beta}
\end{figure}


First, we analyze the goodput characteristics of HDCP for different choices of $\beta$.
In Figure~\ref{fig:beta:1}, we compare the goodputs of each of the forty four nodes for six different choices of $\beta$.\ 
This figure clearly shows that the goodput for $\beta=1$ and $0.8$ are significantly better than the other choices of $\beta$.\
It also demonstrates that $\beta=0$ results in a gradual decrease in the goodput to sink where only few nodes are able to reach the maximum possible rate.
This figure also shows that the choice of $\beta \in \{0,0.2,0.4,0.6\}$ does not significantly affect the goodput performance.
Based on these observations, we hypothesize that the goodput performance of the network is mostly dependent on the ETX of the path and, therefore, for higher $\beta$ values the goodput performance metrics are better.
Figure~\ref{fig:beta:2}, which shows that the average path costs for $\beta \in \{0.8,1\}$ in terms of ETX for individual sources are significantly less than the average path costs for other choices of $\beta$, validates this hypothesis.
In Figure~\ref{fig:beta:2}, we also analyze the average hop counts observed by the packets. It shows a similar pattern as the path costs since total ETX of the path is proportional to the number of hops traversed by the packet.

In Figure~\ref{fig:beta1} (Bottom), we analyze the variation in the average end-to-end delay suffered by the packets generated from individual nodes for different values of $\beta$.
It shows that the average delay performance for $\beta=1$ is the best among different choices of $\beta$ while any other choice of $\beta$ results in a worse delay performance. 
Similar statistics are seen in Figure~\ref{fig:beta1} (Top) in terms of the average queue sizes for individual nodes.
This figure demonstrates that for $\beta = 1$ the average queue-sizes are almost three to four times smaller than that of the average queue sizes for $\beta \in \{0,0.2,0.4,0.6\}$. 
We also plot the delay cdf in Figure~\ref{fig:beta_cdf} for the packets generated from mote 38 in the testbed which is the mote farthest from the sink. It also shows that $\beta=1$ is best in terms of end-to-end delay.

Summarizing all these results, we can say that HDCP  performs really well if the value of $\beta$ is close to $1$. For lower values of $\beta$, we find the performance does not differ by too much from the performance when $\beta = 0$ (the reason for this is further discussed in section~\ref{sec:similarity}).
Therefore, we only consider HDCP with $\beta=1$ and $\beta=0$ (the latter as a baseline scheme, which does not take into account ETX) for the rest of the paper.


\begin{figure}[!ht] 
 \centering
 \subfloat[]{\label{fig:mod} \includegraphics[height=0.4\linewidth,width=0.5\linewidth]{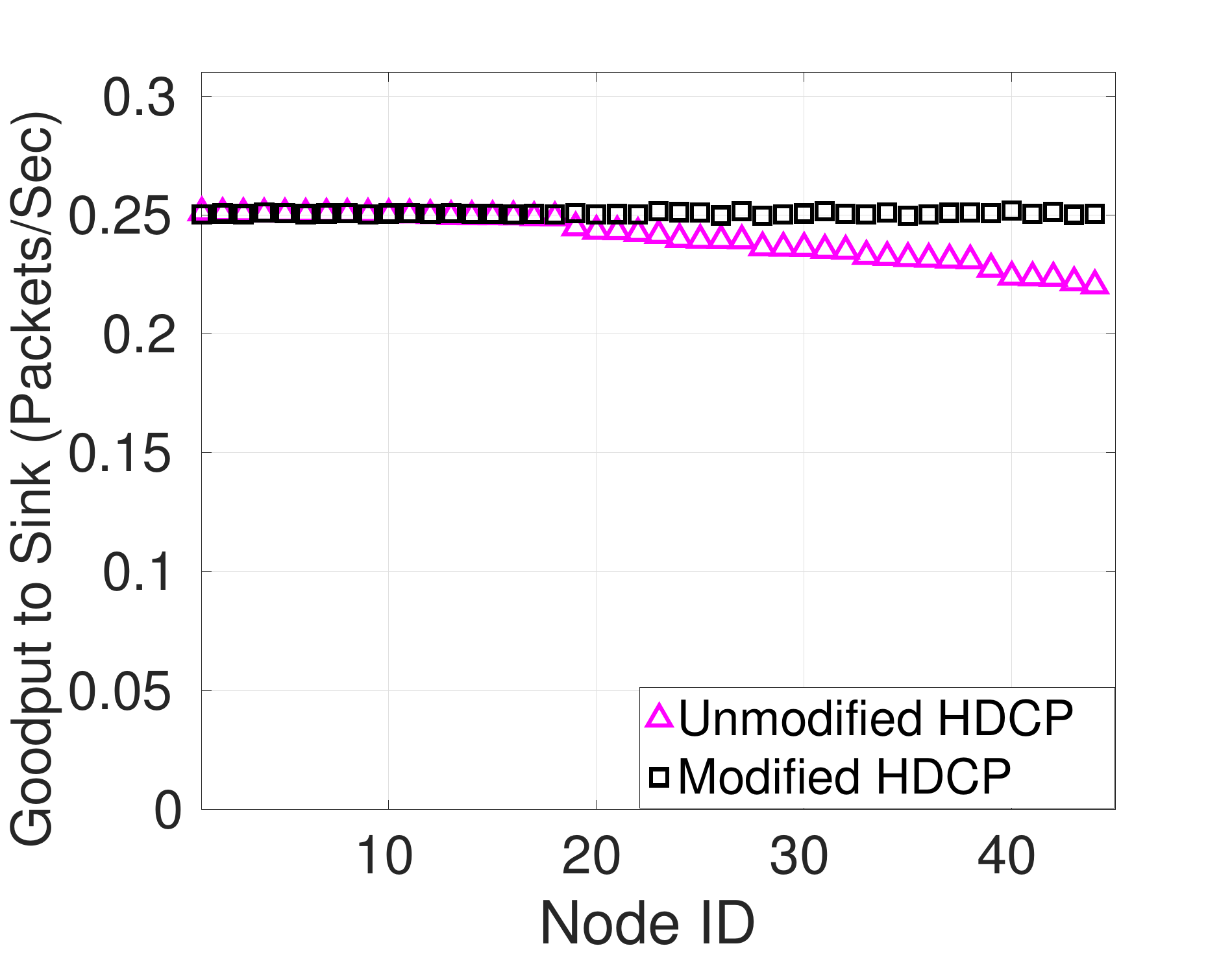}}\,
 \subfloat[]{\label{fig:mod:2} \includegraphics[height=0.5\linewidth,width=0.5\linewidth]{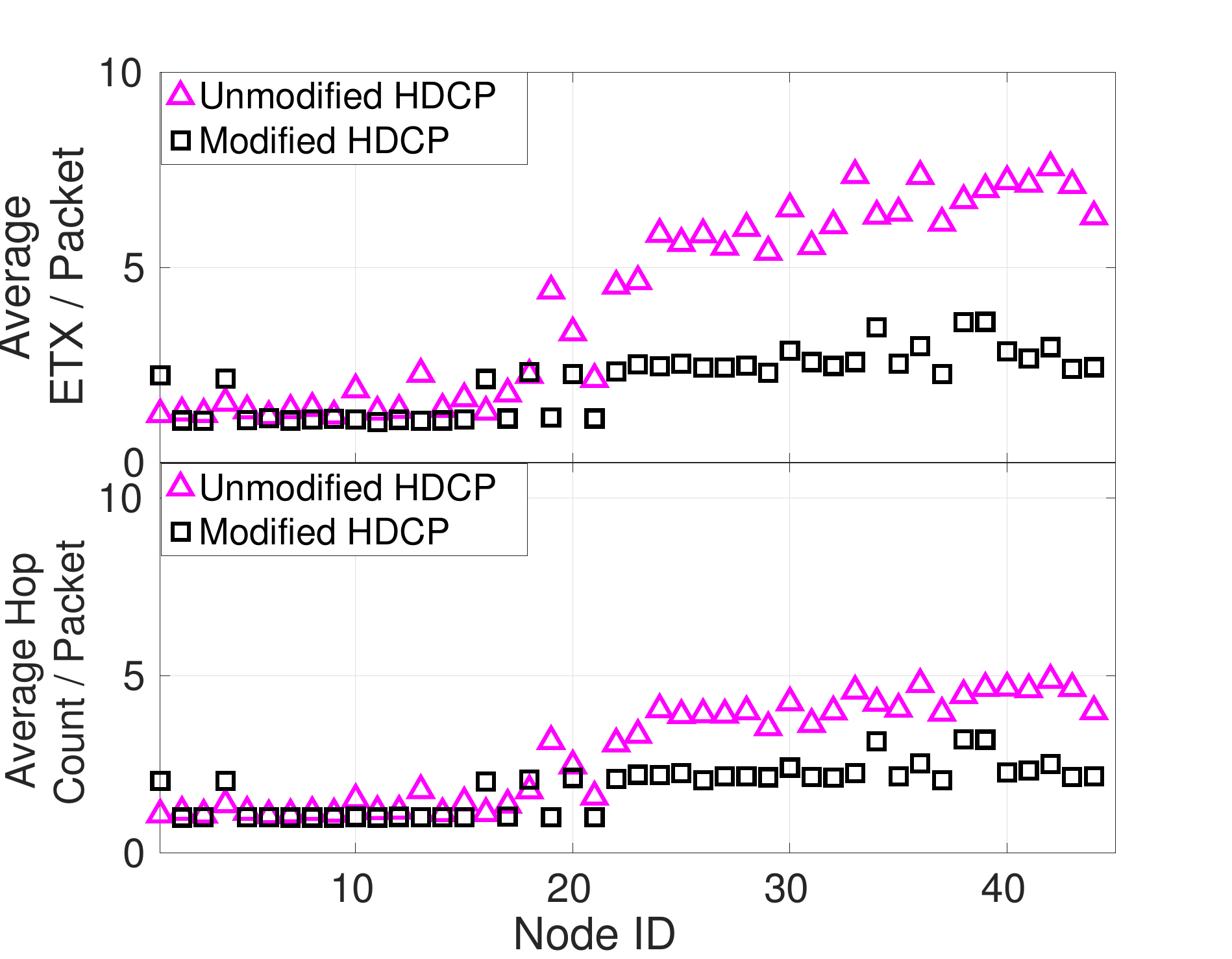}}
 \subfloat[]{\label{fig:mod:4}
 \includegraphics[height=0.5\linewidth,width=0.5\linewidth]{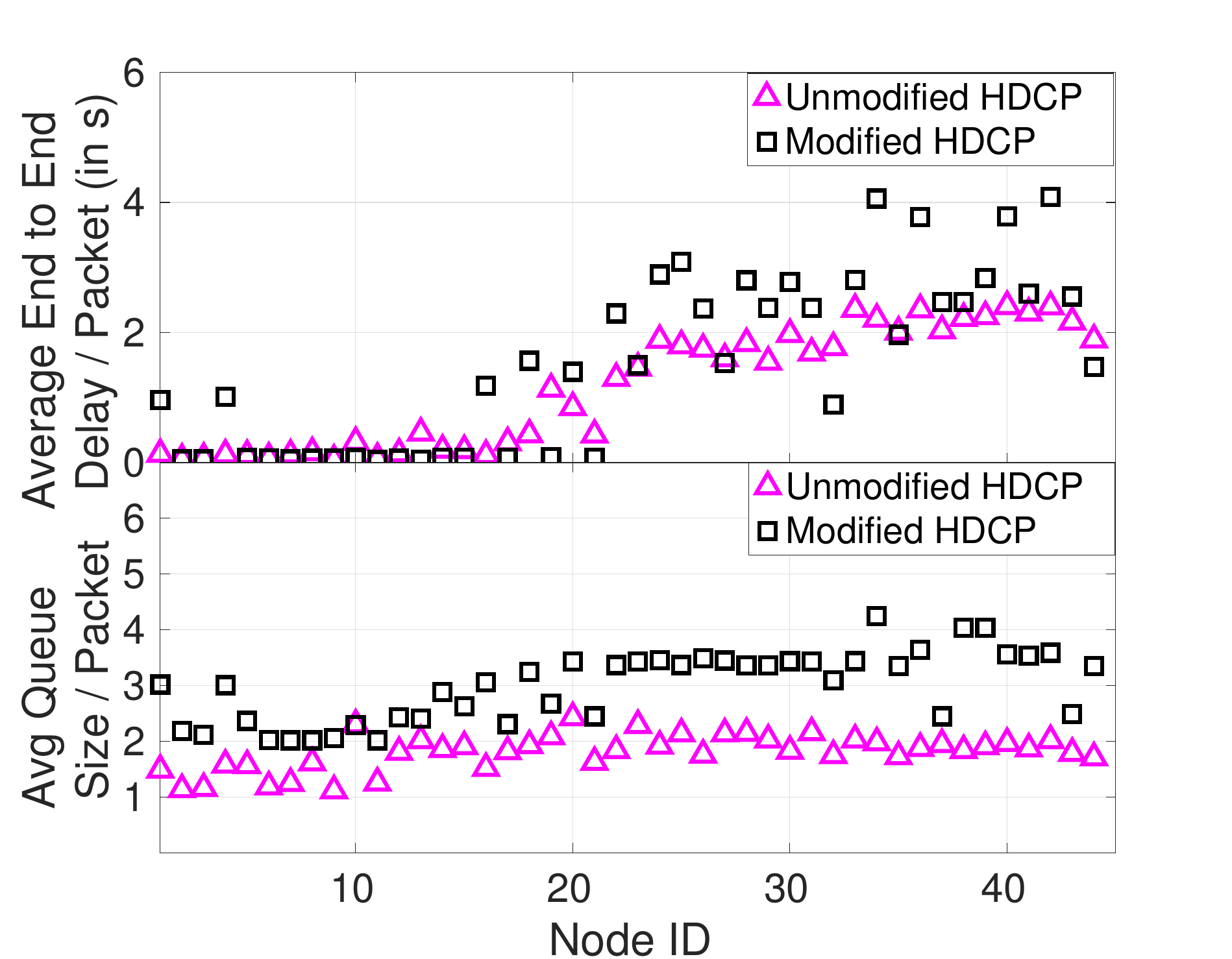}}
 \caption{Performance Comparison between Modified and Unmodified HDCP Implementation with $\beta=1$ for 0.25 PPS: (a) Average Goodput (b) Average ETX per Packet (Top) and Average Hop Count (Bottom) (c) Average End-to-End Delay (Top) and Average Queue Size (Bottom)}
\end{figure}

\subsection{Modified HDCP vs Unmodified HDCP}
\label{sec:unmod}

In this section, we present a comparison of an HDCP implementation based on the original weighing model suggested by the theory (Eqn.~\eqref{eq:hd}) with our HDCP implementation where the link weight model is modified as in~\eqref{eq:hd2_mod1} as well as with our proposed link switching approach. For this purpose, we perform a set of experiments with both versions of HDCP for a fixed value of $\beta=1$ and fixed packet generation rate of 0.25 PPS i.e., 1 packet per 4 seconds.
Note that, for visual clarity, all the plots presented in this section are sorted in terms of the goodputs for unmodified HDCP implementation.

In Figure~\ref{fig:mod}, we demonstrate that without the modifications we have proposed, the goodput performance of HDCP suffers significantly.
This is mostly due to the selection of links with higher ETX as well as lack of proper queue gradient towards the sink, as discussed in Section~\ref{sub:qpara}. 
This is further verified by the Figure~\ref{fig:mod:2} which clearly shows that the average path costs in terms of ETX for unmodified HDCP are very high compared to our HDCP implementation.


%

Next, we compare the performance of unmodified and modified HDCP in terms of average end to end delay as well as average queue size of individual nodes in Figures~\ref{fig:mod:4}. It shows that the delay performance of unmodified HDCP is worse than modified HDCP for half of the nodes while it is better for the rest half of the nodes. Thus, on average, the modification does not attribute to any delay improvements. 
On the other hand, it is also clear from the figure that there exists a steady queue gradient in modified HDCP in contrary to the case of unmodified HDCP, where most of the nodes have queue size of 1 thereby lacking a proper queue gradient towards sink. This also validates our justification for the modification of weights in HDCP as indicated in Section~\ref{sub:qpara}. Thus, overall we improve the performance of HDCP by slightly compromising the average queue sizes.


\subsection{Performance Comparison with BCP and CTP for Fixed Packet Generation Rate}

In this section, we compare the performance of HDCP with the performance of the BCP protocol and the CTP protocol for the fixed packet generation rate of 0.5 PPS, i.e., 1 packet per 2 seconds.
Note that, for simplicity of presentation, all the plots presented in this section are sorted in terms of the goodputs for the BCP algorithm. 
\begin{figure}[!ht]
 \centering
 \subfloat[]{\label{fig:bcp_hd:1} \includegraphics[height=0.4\linewidth,width=0.5\linewidth]{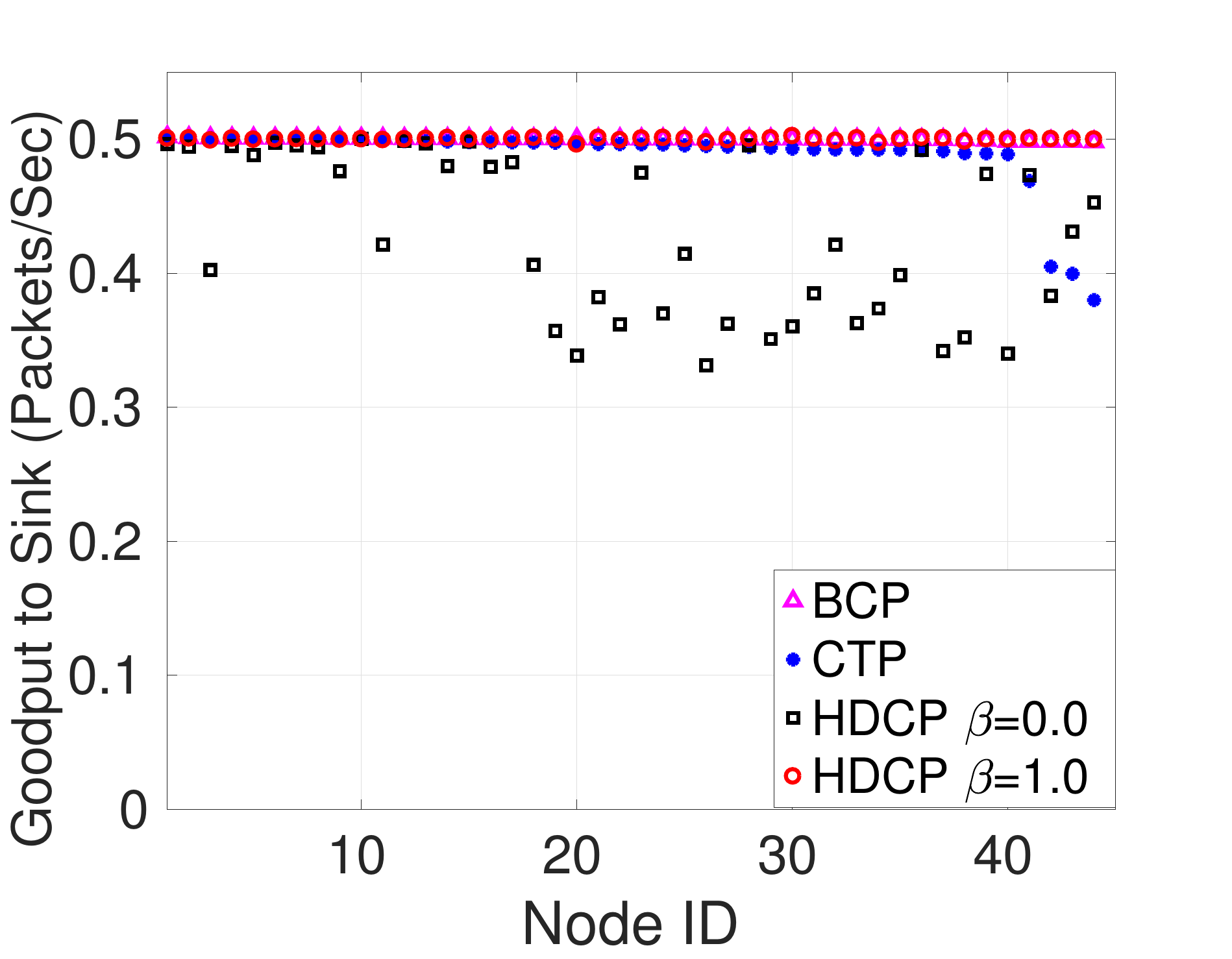}} \,
 \subfloat[]{\label{fig:bcp_hd:2} \includegraphics[height=0.5\linewidth,width=0.5\linewidth]{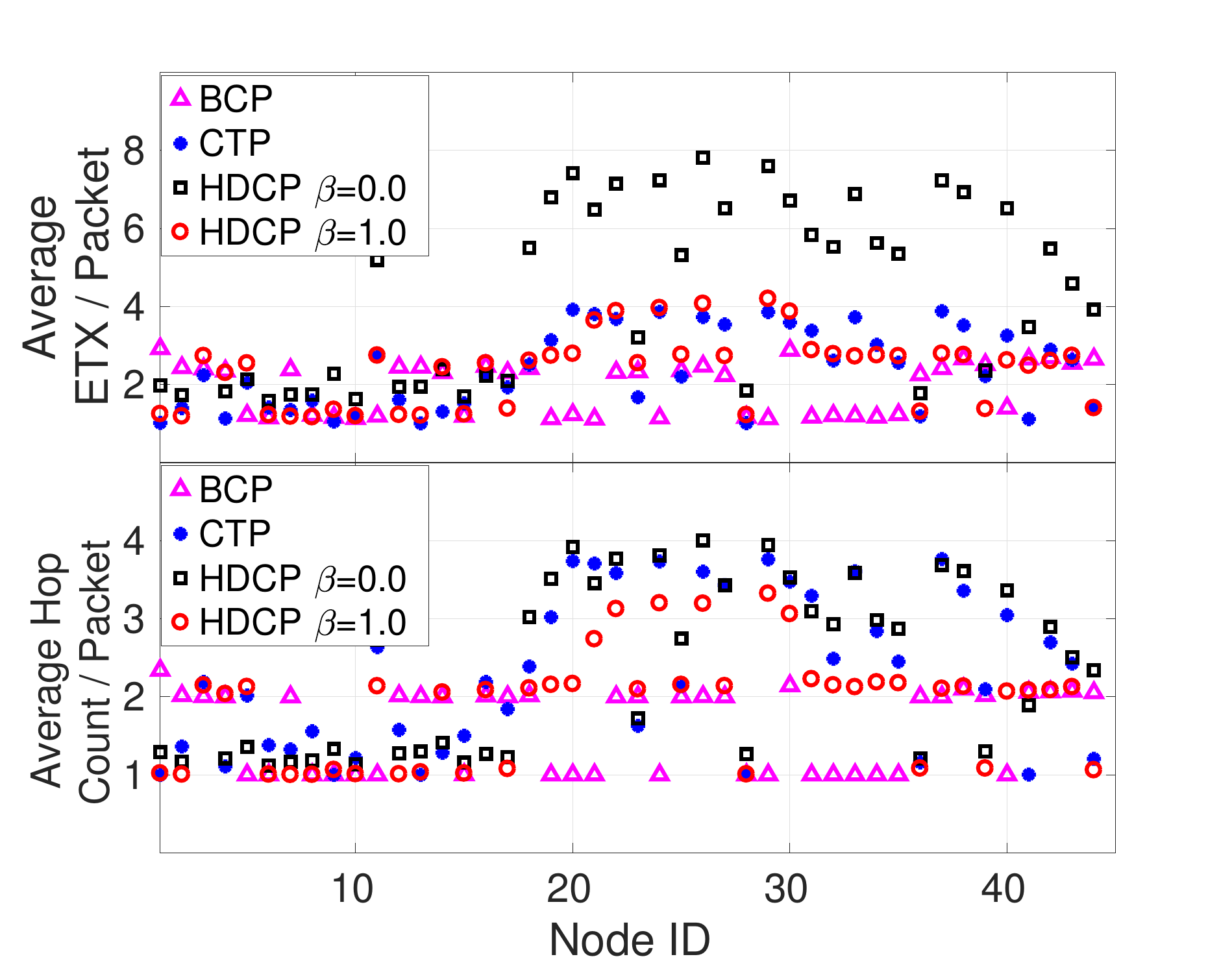}}
 \subfloat[]{\label{fig:bcp_hd1:1}
 \includegraphics[height=0.5\linewidth,width=0.5\linewidth]{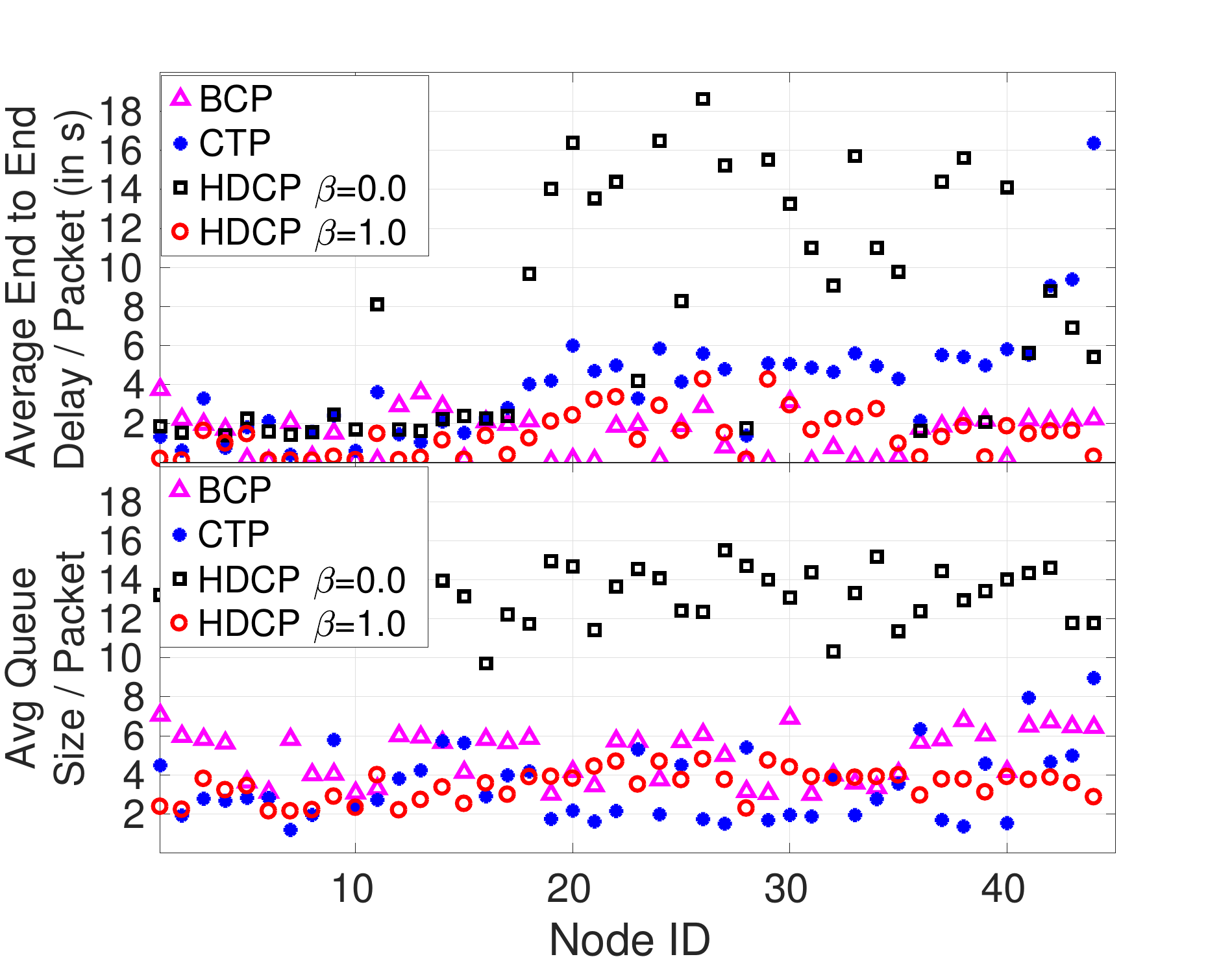}}
  \caption{Comparison Plots between HDCP, BCP and CTP for 0.5 PPS: (a) Average Goodput to Sink (b) Average ETX (Top), Average Hop Count to Sink (Bottom) (c) Average End-to-End Delay (Top) and Average Queue Size (Bottom)}
 \label{fig:bcp_hdd}
\end{figure}

In Figure~\ref{fig:bcp_hd:1}, we plot the goodputs for CTP, BCP and HDCP with $\beta=0$ and $1$, respectively.
We observe that HDCP with $\beta=1$ outperforms the CTP algorithm in terms of goodput while CTP outperforms HDCP for $\beta=0$. However, BCP and HDCP with $\beta=1$ performs almost identically.
For $\beta=0$, the weights of the links are fully determined by the queue differentials and it does not depend on ETX at all, resulting in bad performance. For CTP, a node relies on a single periodically calculated path to sink and does not take advantage of multiple available paths to sink thereby compromising the goodput for high packet generation rate such as 0.5 PPS. On the other hand, BCP and HDCP with $\beta=1$ focus on reducing the total ETX cost of a source to sink path while not being restricted to a single pre-calculated path. Thus, the BCP and the HDCP algorithm with $\beta=1$ appear to be able to take advantage of the multiple paths available to the sink in order to cope with high packet generation rates thereby improving the throughput region.

Similar to the goodput analysis, we present the average hop count and average ETX of the entire path observed by the packets generated from individual sources in Figure~\ref{fig:bcp_hd:2}.\
It shows that, again, HDCP with $\beta=1$ and BCP both slightly outperform CTP on average whereas HDCP with $\beta=0$ performs the worst. 
This is also justified based on our discussion presented in the previous section. 
The performance of HDCP with $\beta=1$ and the performance of BCP are again almost same.
\textbf{Based on these results, we hypothesize that the similarity between BCP and HDCP with $\beta=1$ is due to similaritsy in their neighbor rankings, despite differences in the structure of the weight expression.}
This is further explored in section~\ref{sec:similarity}.



We also compare the delay performance and queue size of HDCP with BCP and CTP in Figure~\ref{fig:bcp_hd1:1}.
Figure~\ref{fig:bcp_hd1:1} shows that the delay performance of HDCP for $\beta=1$ is significantly better than HDCP with $\beta=0$.
However, based on the figure, the delay performance for BCP is almost same as HDCP with $\beta=1$ while both of them outperforms CTP. 
The similarity between BCP and HDCP with $\beta=1$ is justified based on the previous results. 
In Figure~\ref{fig:bcp_hd1:1}, we also demonstrate that the average queue size of HDCP with $\beta=1$ is significantly low compared to BCP and HDCP with $\beta=0$. 
The queue size of CTP seems to be the lowest for some nodes, however, we believe this is misleading as CTP experiences the most packet drops among the various protocols at this offered load. The packet drops in CTP occur partly due to retransmission packet drops caused by higher intra-network interference (reflected in the higher ETX and higher delay values), and partly due to some other parameters in its implementation such as forwarding packet lifetime and an in-built congestion control. However, for any higher packet generation rate, we observe that the queue size for CTP increases rapidly (resulting in even more losses) as does its delay.



\subsection{Varying Packet Generation Rate}
\begin{figure}[!ht]
    \centering
    \subfloat[]{\label{fig:offer_goodput}   \includegraphics[height=0.4\linewidth,width=0.5\linewidth]{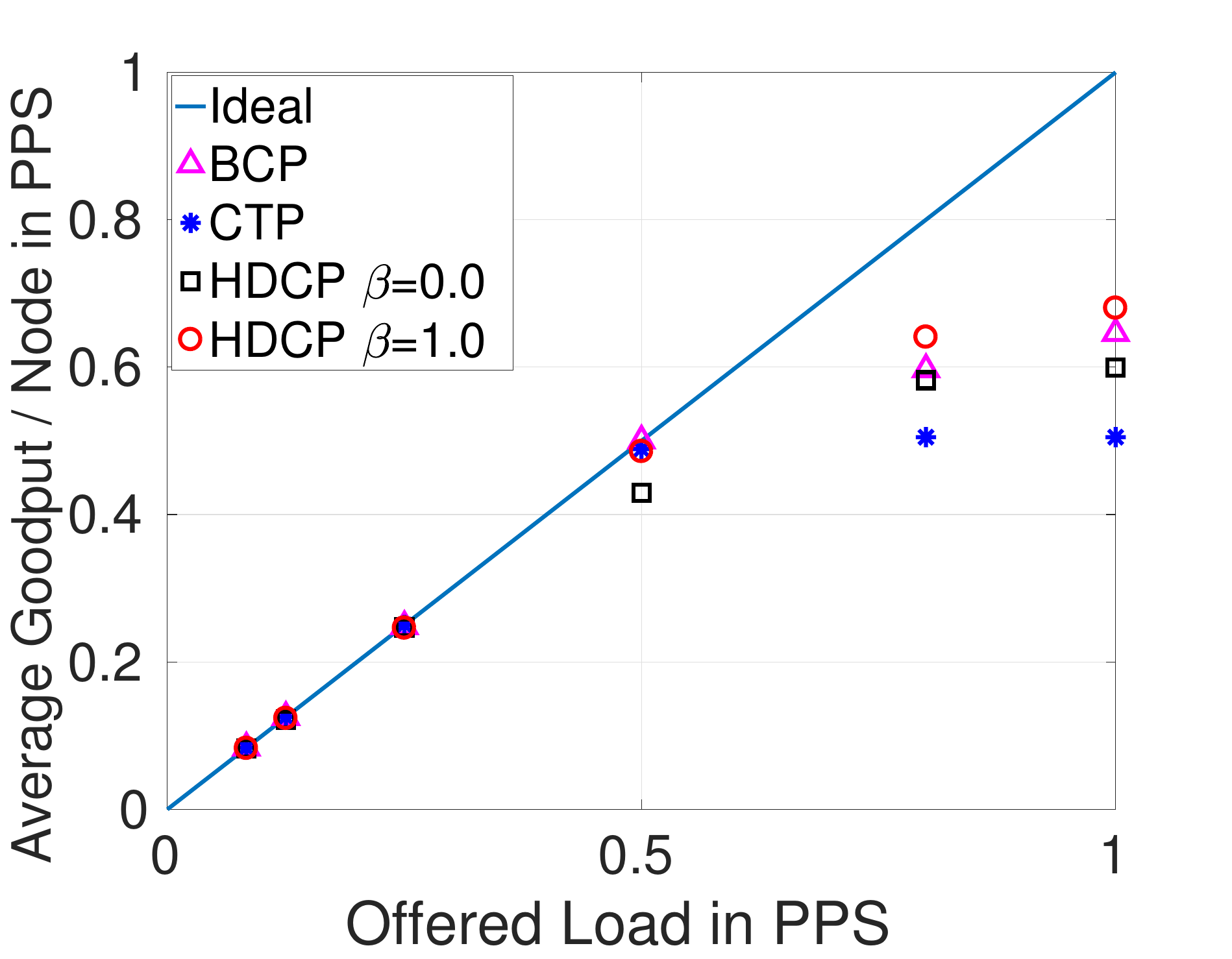}}
    \subfloat[]{\label{fig:delay} \includegraphics[height=0.4\linewidth,width=0.5\linewidth]{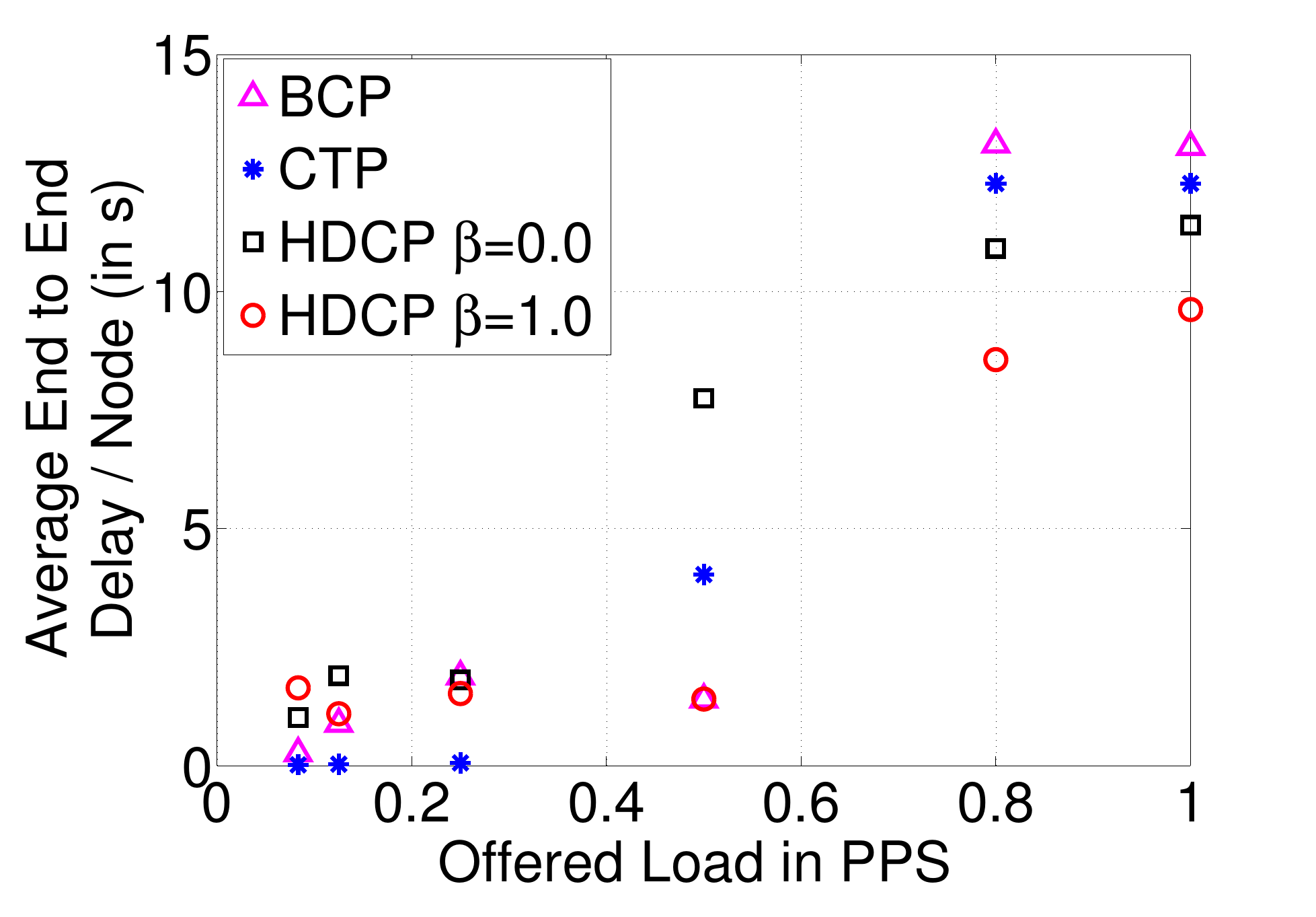}}\,
    \subfloat[]{\label{fig:offer_etx} \includegraphics[width=0.6\linewidth]{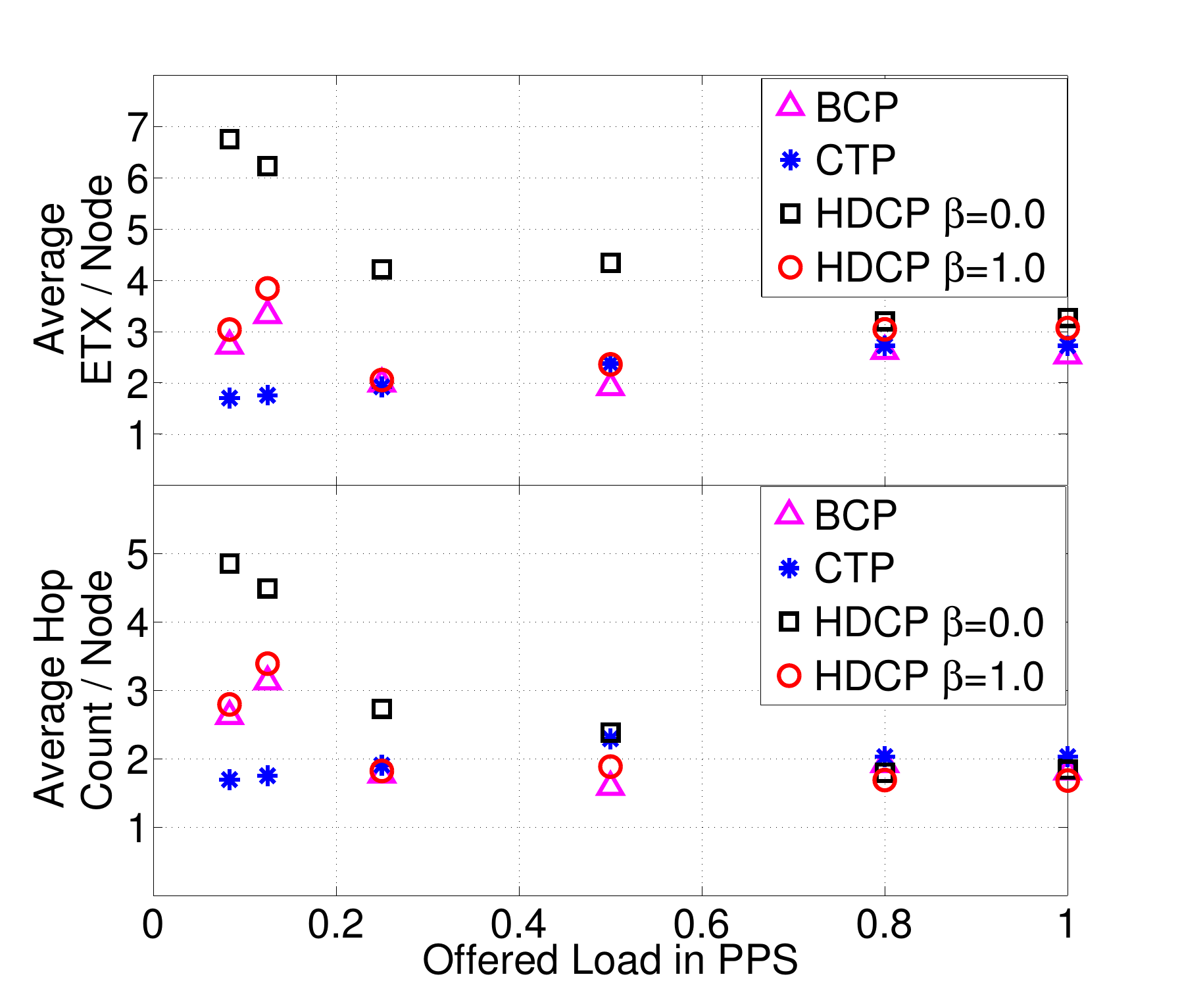}}
   \caption{(a)Variation of Goodput for Varying Offered Load (b) Variation of Average End to End Delay for Varying Offered Load (c) Variation of Average Path Cost in Terms of ETX (Top) and Average Hop Count (Bottom) for Varying Offered Load}
\end{figure}

In this section, we present and analyze the effects of the packet generation/source rates on the performance of HDCP and compare it with the performance of the BCP and CTP algorithms.\
We performed a set of experiments with six different packet generation rates: 1/12 PPS (i.e., 1 packet per 12 second), 1/8 PPS, 1/4 PPS, 1/2 PPS, 4/5 PPS, and 1 PPS.
In Figure~\ref{fig:offer_goodput}, we present the goodput variation due to the change in packet generation rate for HDCP with $\beta=0$ and $1$ as well as the goodput variations of the BCP and the CTP algorithms.\
It is clear from Figure~\ref{fig:offer_goodput} that for lower packet generation/source rates, HDCP performs almost similar to the BCP and CTP algorithm in terms of goodput to sink.
But, as we increase the offered load, HDCP and BCP gradually outperform the CTP algorithm.
In our experiments, HDCP outperforms CTP in terms of goodput for packet generation rates higher than 1 packet per 4 seconds.
From the figure, we can estimate that the full throughput region (the maximum offered load at which the protocol is able to match the ideal curve) for HDCP is about 60 to 100\% higher than that for CTP in this particular testbed and topology (of course the relative performance improvement is certainly likely to depend on the network topology.) Another thing to notice that, the average goodput for $\beta=1$ is always higher than $\beta=0$ which agrees with our earlier findings and arguments concerning the inefficiencies introduced by ignoring the ETX costs of links. Yet again, the performance of BCP closely follows the performance of HDCP with $\beta=1$ which is, again, due to similarity in their neighbor rankings in terms of the weights.
This is further explained in section~\ref{sec:similarity}.

Next, we investigate the effects of increasing packet generation rates on the average path costs in terms of ETX and the average number of hops traversed by the packets.\
We plot the average ETX and average hop counts due to different source rates for HDCP, BCP and CTP in Figure~\ref{fig:offer_etx}.\
It is observable from the figure that for any packet generation rate overall path cost for HDCP with $\beta=1$ is comparable to BCP while CTP outperforms both for packet generation rate lower than 0.25PPS and converges with them for higher rates.
Moreover, the average path cost for HDCP with $\beta=0$ is higher than $\beta=1$ which is justified by our discussion in the previous section.
Similar statistics is available from the plot of average numbers of hops encountered by each packet due to its direct relation with the overall path ETX.
The similarity between HDCP with $\beta=1$ and BCP is, again, justified based on our earlier discussions.
The apparent `good' performance of CTP is due to its increasing incapability of sending packets with long path costs to the sink as it encounters congestion drops.

Lastly, we analyze the effect of packet generation rate on the average delay in Figure~\ref{fig:delay}.
Although CTP and BCP outperform HDCP for source rates lower than 0.25PPS by a small margin, this figure demonstrate the superiority of the HDCP for $\beta=1$ in overall delay performance as it continues to guarantee lower delay for higher packet generation rates.
Another interesting fact to notice is that for BCP and HDCP, the delay gradually increases with packet generation rate whereas the delay for CTP increases rapidly with packet generation rate. This is likely because BCP and HDCP can take advantage of multiple paths to sink whereas the CTP relies on only one path. Therefore, CTP reaches congestion earlier than BCP and HDCP which results in the rapid increase in delay.
Again, the similarity between BCP and HDCP with $\beta=1$ is due to similarity in the neighbor ranking in terms of the weights. 


To summarize, our experiments lead us to conclude that optimized combinations of queue-awareness and ETX (implemented in BCP and HDCP with $\beta = 1$) provide the best choice for routing, better than routing based on ETX alone (CTP), which in turn performs better than queue-aware routing alone (HDCP with $\beta = 0$).

\subsection{Low Power Communication Stack Based Experiments}
\label{sec:lowpower}

\begin{figure}[!ht]
    \centering
   \includegraphics[width=0.7\linewidth]{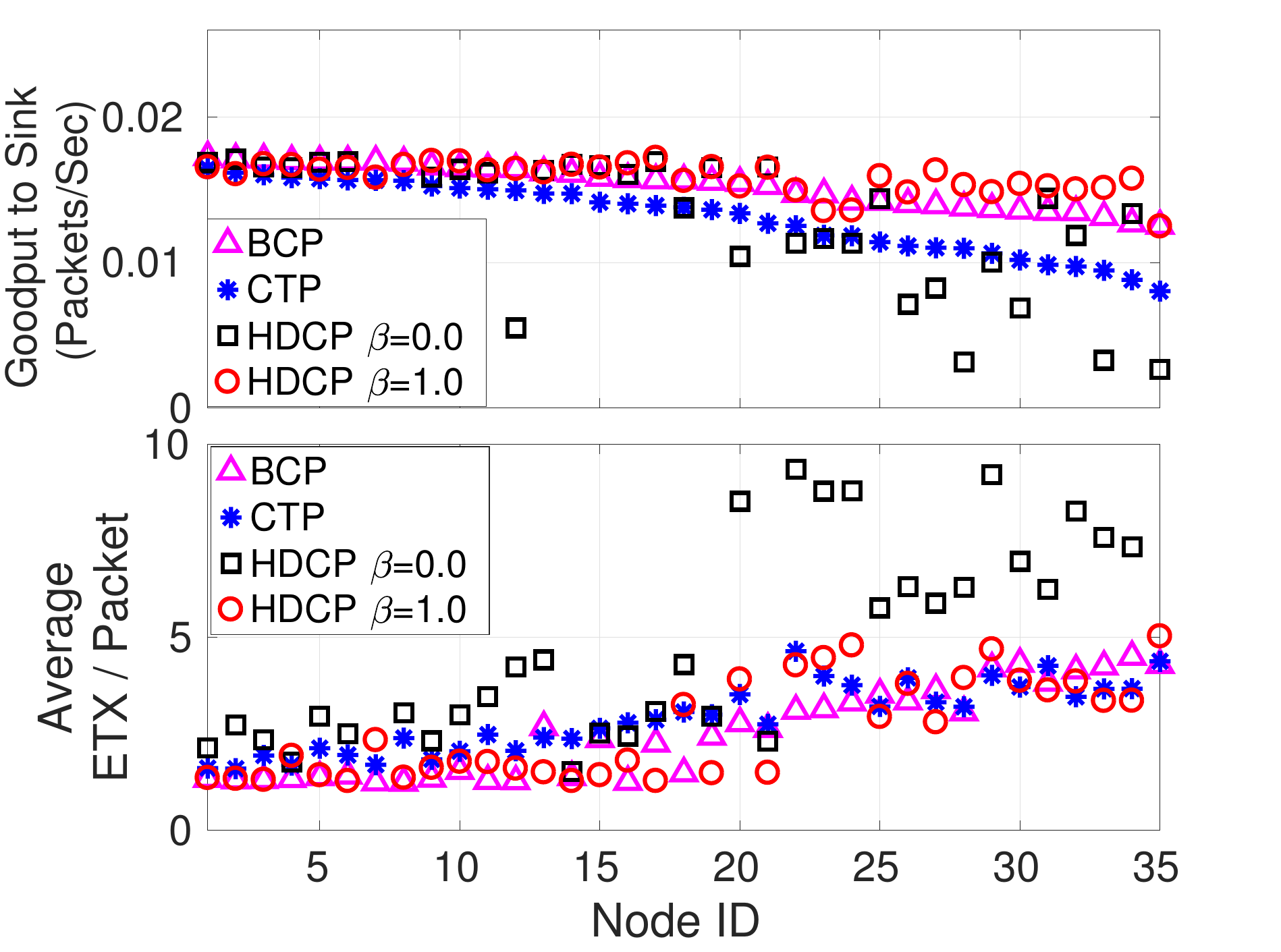} 
   \caption{Performance Comparison of HDCP with BCP and CTP for a Low Power Communication Stack: (Top) Goodput to Sink, (Bottom) Average ETX Path Costs to Sink}
    \label{fig:lowpower}
\end{figure}

In order to verify the performance of HDCP on a low power communication stack, we performed a set of experiments with 35 sources and a sink (first 36 nodes of the testbed). For these experiments, we used CX-MAC protocol, a version of X-MAC~\cite{buettner2006x} that is provided in Contiki, with duty cycle of $5\%$ for HDCP, BCP and CTP. However, the choice of CX-MAC protocol over the other protocols is just a matter of the availability of Contiki implementation.
Furthermore, since we are using a duty cycle, we also need to cut-back our source rates to a very low rate. For the presented set of experiments, we used a packet generation rate of 1 packet per 60 seconds (i.e., 1/60 PPS). We present the results in Figure~\ref{fig:lowpower}. This figure shows that the HDCP protocol with $\beta = 1$ performs well in a low power communication stack, at a very low duty cycle setting where even CTP shows some deterioration in fairness of goodput. However, in this setting the performance of the baseline with $\beta = 0$ is much worse, leading us to conclude that it is a very poor setting indeed. \todoo{Now, in order to estimate the actual energy consumptions, we record the different energy consumption components using the Contiki PowerTrace tool (in terms of the percentage of time spent in different radio phases: Transmit, Listen/receive). Based on our traces, in HDCP with 5\% duty cycle, the radio of each node is on for $5.92\%$ of the total execution time, out of which the node is transmitting and receiving approximately $0.65\%$ and $5.27\%$ of the total execution time, respectively. Now, to get the actual energy consumption, one can use the current and voltage ratings from the specifications of the devices used. For example, in Tmote-sky the rated voltage of operation is approx $3.3V$ and the average current consumptions are $17.4mA$ and  $19.7mA$ for radio transmission and radio reception, respectively. This results in approximately $113.78mJ$ energy consumption in each Tmote-Sky for the experiment period of 30 minutes.}

\subsection{External Interference}
\label{sec:interf}
\begin{figure}[!ht]
 \centering
 \subfloat[]{\label{fig:inter}\includegraphics[height=0.4\linewidth, width=0.5\linewidth]{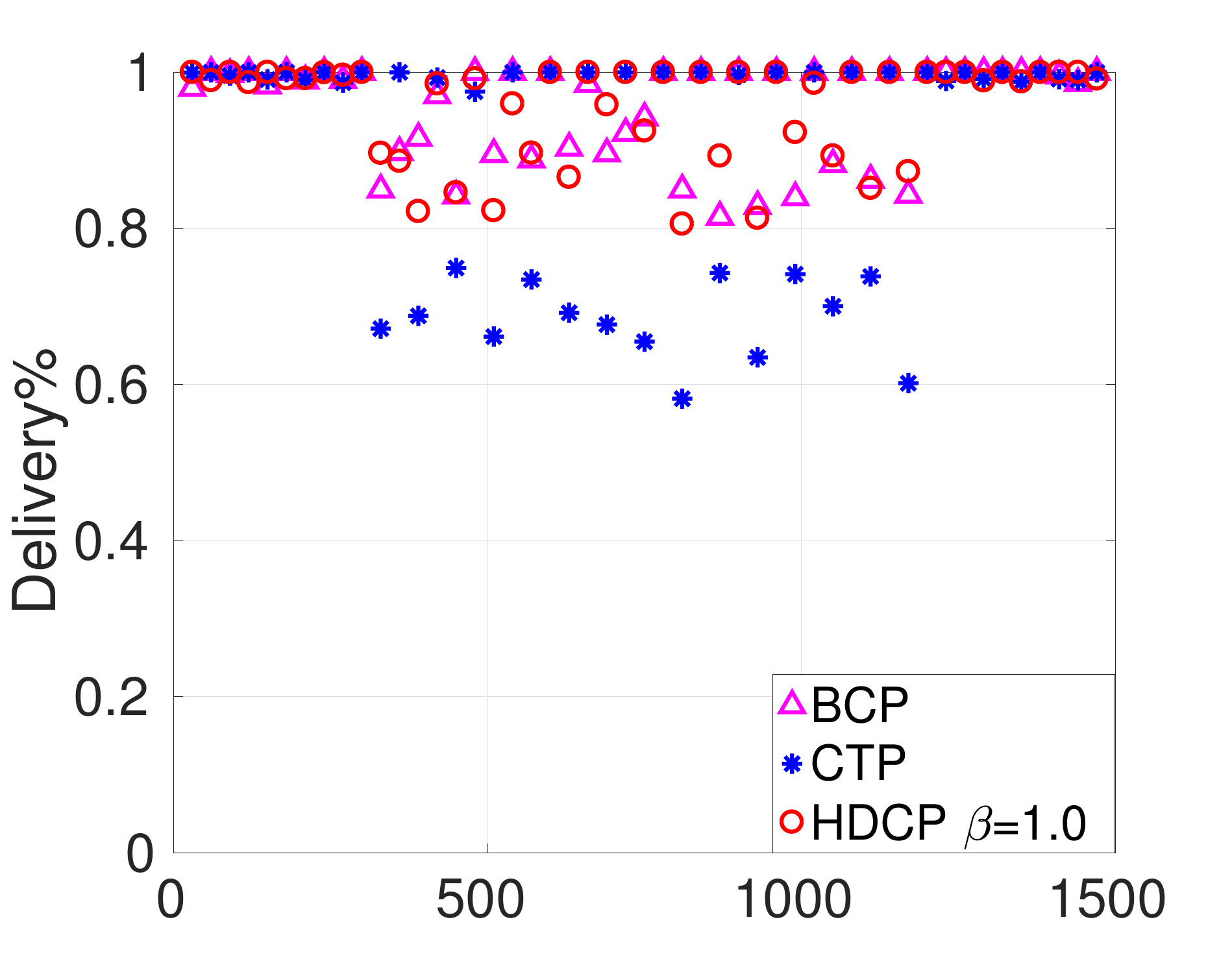}}
  \subfloat[]{\label{fig:inter1}\includegraphics[height=0.4\linewidth, width=0.5\linewidth]{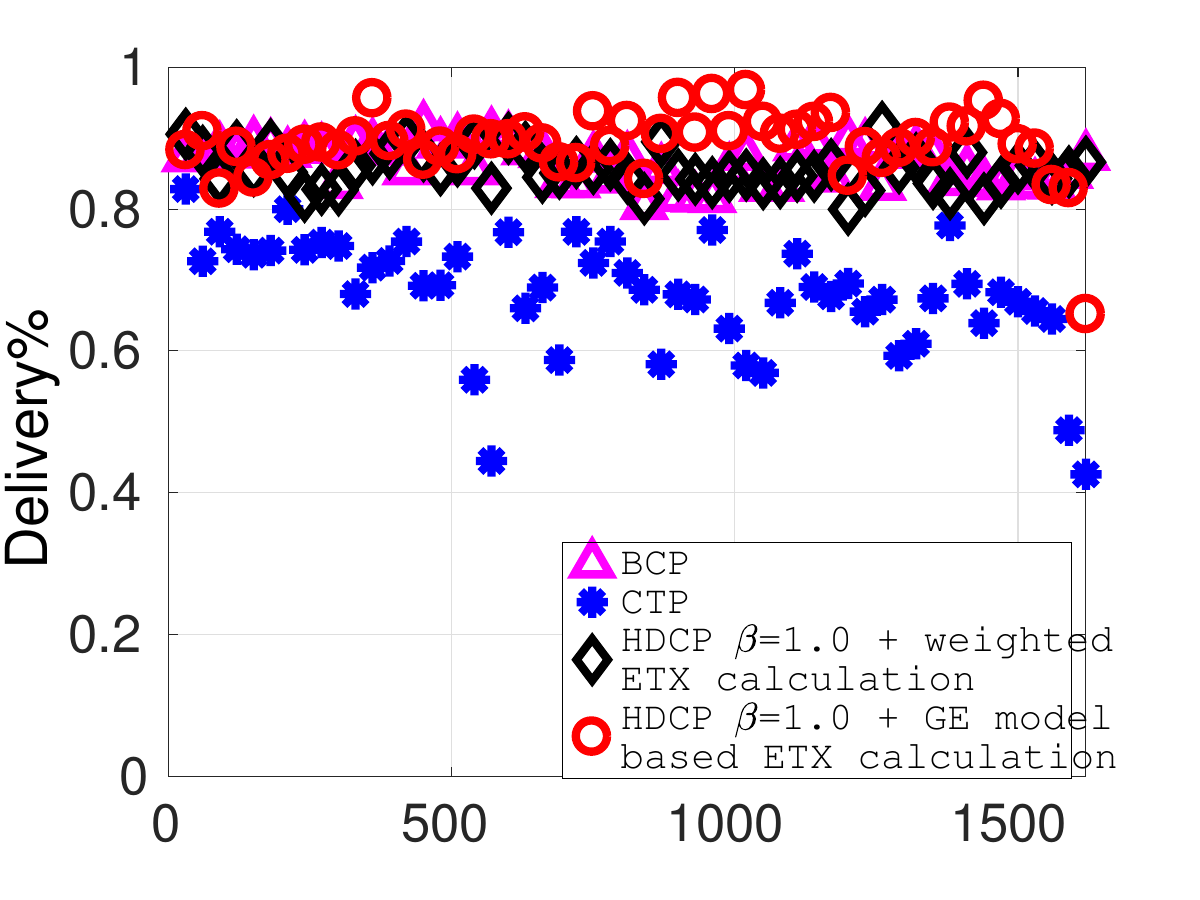}}
  \caption{Thirty Second Windowed Average Sourced Packet Delivery Ratio for: (a) Synthetically Generated Interfering 802.15.4 Channel 26 Traffic (b) Real Interference Scenario on 802.15.4 Channel 13}
 \label{fig:interfere}
\end{figure}

In this section, we evaluate the performance of the HDCP protocol with the optimized $\beta = 1$ in the presence of external interference and compare it with both BCP and CTP. This is necessary because the 802.15.4 radios share frequency band with WiFi, Bluetooth, and other Zigbee radios and as a result their performance often suffers from severe interference. To emulate such scenarios, we performed a set of experiments with forty sources and a single sink (Node 1) while four nodes are used as interference sources on channel 26. 
The interfering nodes are inactive for the first five minutes of the experiment, periodically transmit for next fifteen minutes, and become inactive again for the last five minutes of the experiment. During the on period, each of the interfering nodes transmits 110 Byte packets at a rate of 100PPS for 15 seconds and then does not transmit anything for the next 15 seconds, and so on. Furthermore, we reduced the power level of all 41 nodes from level 31 to level 15 whereas the interfering nodes were kept at level 31, in order to intensify the effect of interference. The outcome of this set of experiments is presented in Figure~\ref{fig:inter} that plots the delivery percentage of the packets over a series of 30 seconds time window for HDCP with $\beta=1$, BCP and CTP. It demonstrates that while CTP performance significantly suffers from the interference, the HDCP protocol maintains its good packet delivery ratio, similar to BCP.

\todoo{The above mentioned settings is used to stay consistent with the interference settings presented in the original BCP paper\cite{moeller2010routing}. 
However, it is well known that the simple Gilbert-Eliot model used for ETX estimation might work perfectly with some specific synthetic interference models and might fail in realistic interference scenarios. In order to explore the performance of the HDCP algorithm, in presence of real interference, we performed a set of experiments with 44 source nodes and 1 sink node, running on channel 13 of the 802.15.4 standard which is known to be one of the most interfered channels. We also compare the performance of HDCP based on the Gilbert-Eliot (GE) ETX model with the performance of BCP with the GE model as well as with HDCP based on the ETX model used in the original BCP paper~\cite{moeller2010routing}. 
The results presented in Figure~\ref{fig:inter1} clearly demonstrate that even in the presence of constant real interference, the HDCP algorithm with Gilbert-Eliot ETX model performs comparable to the BCP algorithm and the HDCP algorithm with the basic ETX model presented in \cite{moeller2010routing}, while outperforms the CTP algorithm. Furthermore, both Figures~\ref{fig:inter} and~\ref{fig:inter1} show that the BCP and the HDCP algorithms can achieve approx $85\%$ delivery ratio in presence of interference while the CTP achieves approx $70\%$.}

\subsection{Node Failures}
\label{sec:failure}
\begin{figure}[!ht]
 \centering
\includegraphics[height=0.4\linewidth, width=0.5\linewidth]{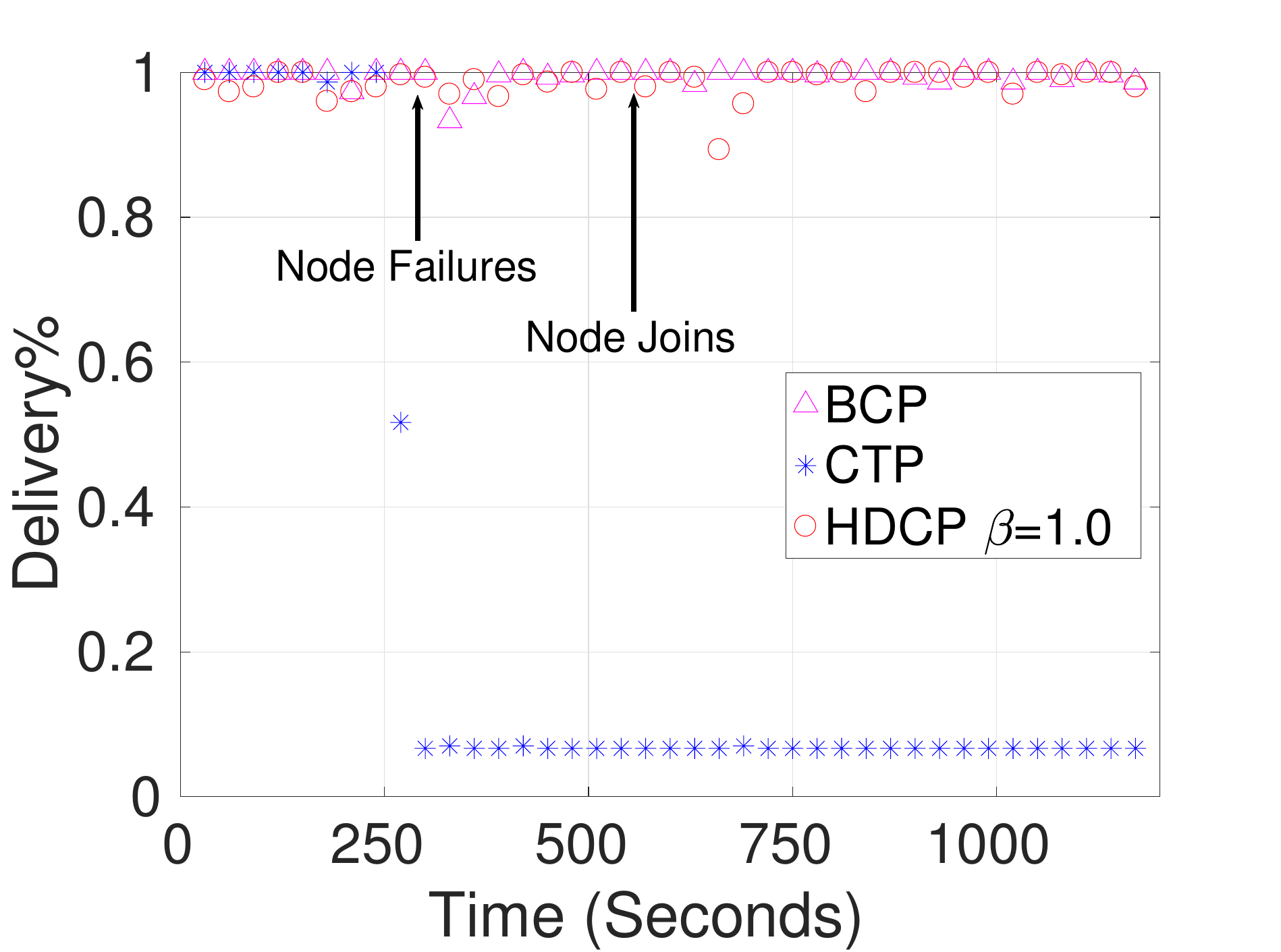}
  \caption{Thirty Second Windowed Average Sourced Packet Delivery Ratio for $10\%$ Node Failures}
 \label{fig:failures}
\end{figure}

In this section, we evaluate the performance of the HDCP protocol with the optimized $\beta = 1$ in the presence of node failures/joins and compare it with both BCP and CTP. 
This is necessary because node failures and node joins are very common events in traditional wireless sensor networks.
To emulate such scenarios, we performed a set of experiments with twenty sources, single sink, and twenty five forwarding nodes, i.e., total forty six nodes in the network. 
All nodes were set to transmit at the maximum power level, i.e., level 31 and on channel 26.  In our experiments, we randomly turned off four of the forwarding nodes (i.e., $\approx 10\%$ nodes) after five minutes from the beginning and then turn them back on after ten minutes from the beginning. 
Each source node was set to transmit at 1/2 PPS.   
A sample outcome of this set of experiments is presented in Figure~\ref{fig:failures} which plots the delivery percentage of the packets over a series of 30 seconds time windows for HDCP with $\beta=1$, BCP, and CTP. 
Figure~\ref{fig:failures} demonstrates that CTP performance significantly suffers after the node failures and could not recover from that due to very high packet generation rate and the reliance on a single predetermined path from each source to the sink.
On the other hand, the performance of both the HDCP protocol and the BCP protocol are unaffected by the node failure/join events. 
This pertains to the fact that both HDCP and BCP do not rely on a single path and, thus, able to take advantage of the alternate paths in the network, after the node failures. This validates that queue-aware routing algorithms such as BCP and HDCP perform well in presence of high node dynamics while the predetermined route based algorithms such as CTP suffer after node failures.
The similarity in performance between BCP and HDCP is again due to similarity in the neighbor rankings in terms of the weights (explained in Section~\ref{sec:similarity}).

\section{Similarity Analysis Between HDCP and BCP}
\label{sec:similarity}

In this section, we analyze the BCP and the HDCP algorithms to identify the reasons behind the similarity in their performance. The performance of both the BCP and the HDCP depend on the rankings of the neighbors (based on the link weighing functions) of a node, which in turn translates to selection of routing paths to sink. 
From theoretical standpoint, the performance of HDCP and BCP will be different in a network if their respective rankings of the neighbors under same queue size conditions are different. \textbf{Conversely, we hypothesize that the similar rankings of neighbors for both the HDCP and the BCP protocol will result in similar performance. }

\subsection{Theoretical Analysis}

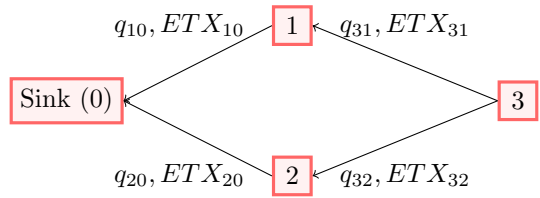
\begin{figure}[!ht]
\centering
\begin{tikzpicture}[
roundnode/.style={circle, draw=green!60, fill=green!5, very thick, minimum size=7mm},
squarednode/.style={rectangle, draw=red!60, fill=red!5, very thick, minimum size=5mm},
]
\node[squarednode]      (source1)    at  (0,1)  {1};
\node[squarednode]      (sink)   at  (-3,0) {Sink (0)};
\node[squarednode]      (source3)  at  (3,0) {3};
\node[squarednode]      (source2)  at  (0,-1) {2};

\draw[->] (source1.west) -- (sink.east);
\draw[->] (source2.west) -- (sink.east);
\draw[->] (source3.west) -- (source1.east);
\draw[->] (source3.west) -- (source2.east);

\node at (-1.5,1) {$q_{10},ETX_{10}$};
\node at (1.5,1) {$q_{31},ETX_{31}$};
\node at (-1.5,-1) {$q_{20},ETX_{20}$};
\node at (1.5,-1) {$q_{32},ETX_{32}$};
\end{tikzpicture}
\caption{A Simple Topology For Ranking Similarity Analysis Between HDCP and BCP}
\label{fig:toy_top}
\end{figure}

\begin{figure}[ht]
    \centering
    \subfloat[]{\label{fig:etx_stat}\includegraphics[height=0.4\linewidth, width=0.5\linewidth]{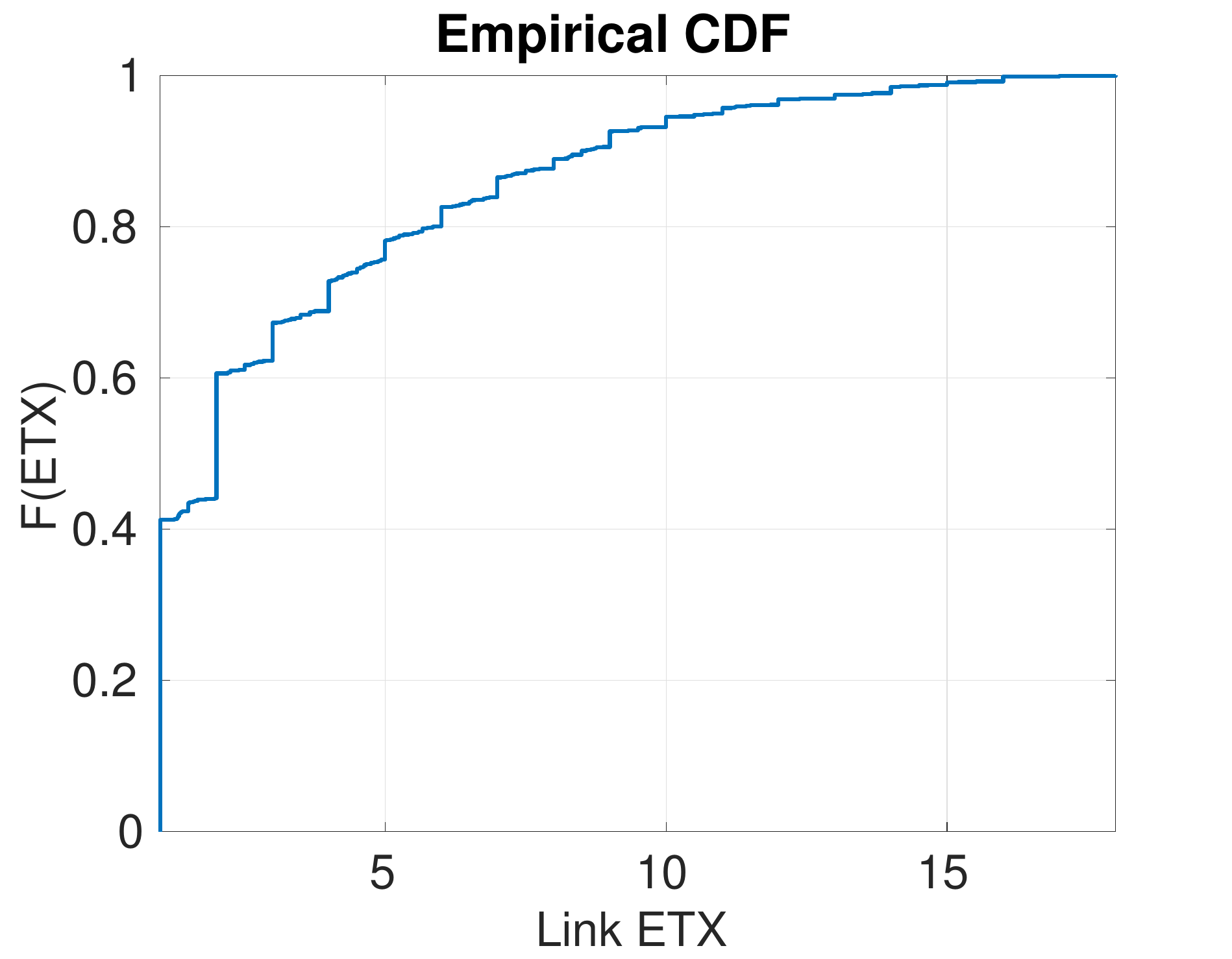}}
    \subfloat[]{\label{fig:indiff_curves}\includegraphics[height=0.4\linewidth, width=0.5\linewidth]{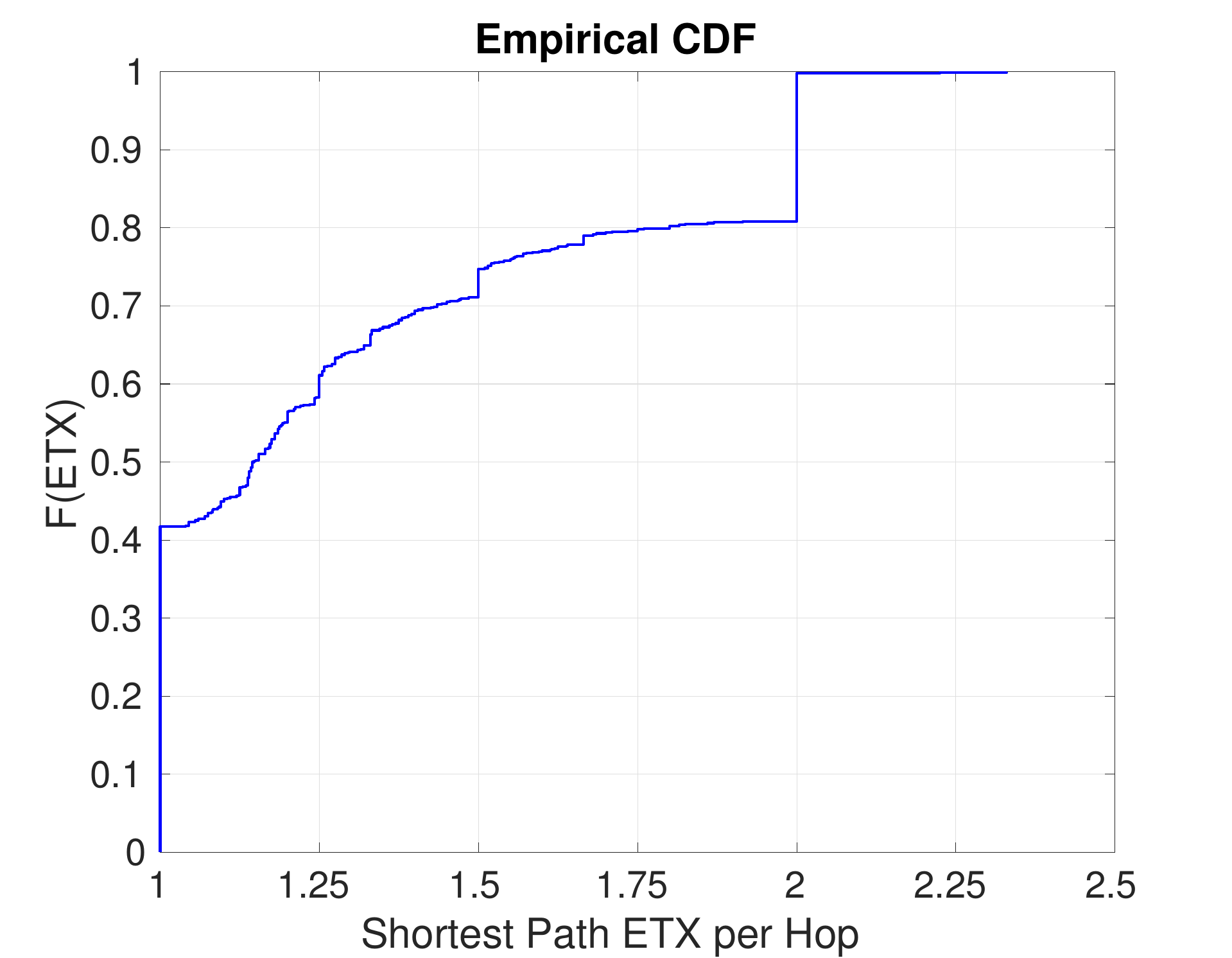}}
   \caption{(a) Empirical CDF of the Link ETX Values for Our Testbed (b) Empirical CDF of the Average ETX per Link for the Shortest Paths Between Any Pair of Nodes}
   
\end{figure}

In order to analyze the scenarios that will result in different or similar rankings of neighbors for HDCP and BCP, we compare the simplified weighing functions of BCP and HDCP with $\beta=1$, which can be written as follows:
\begin{equation}
\label{eqn:compare}
\begin{split}
&w^{bcp}_{ij}(n)= q_{ij}(n) - 2 .\overline{ETX_{ij}}(n)\\
&w^{hdcp}_{ij}(n) = \frac{q_{ij}(n)-\overline{ETX_{ij}}(n)}{\overline{ETX_{ij}}(n)}
\end{split}
\end{equation}
provided that $V=2$ and $\frac{q_{ij}(n)}{2\overline{ETX_{ij}}(n)}\geq 1$, i.e., the links have non zero weights according to both BCP and HDCP weighing schemes. 
First of all, we try to identify the range of the possible network configurations that will result in different rankings for HDCP and BCP. For this purpose, we analyze a toy topology illustrated in Figure~\ref{fig:toy_top}.
Assume that the $ETX_{31}$ and $ETX_{32}$ are $1$ and $e\geq 1$, respectively. Now, the weights of the respective links according to BCP will be:
\begin{equation}
\begin{split}
w^{BCP}_{31}=q_{31}-2 \qquad \mbox{and} \qquad  w^{BCP}_{32}=q_{32}-2e\
\end{split}
\end{equation}
Similarly, the weights for the links according to the HDCP rule for $\beta=1$ will be (provided that $q_{31}\geq 2$ and $q_{32}\geq 2e$):
\begin{equation}
\begin{split}
w^{HDCP}_{31}={q_{31}}-1 \qquad \mbox{and} \qquad w^{HDCP}_{32}=\frac{q_{32}}{e} -1\
\end{split}
\end{equation}
Now, 
\begin{equation}
\label{eqn:toy}
\begin{split}
&w^{BCP}_{31} > w^{BCP}_{32}\ \ if \ \ e > (q_{32}-q_{31})/2 +1\\
&w^{HDCP}_{31} > w^{HDCP}_{32}\ \ if \ \ e > \frac{q_{32}}{q_{31}}
\end{split}
\end{equation}
Thus, if $ \frac{q_{32}}{q_{31}} < e < (q_{32}-q_{31})/2 +1$, the rankings of the outgoing links of node $3$ are different, while the rankings are the same for all other values of $e$. As an example, say, $q_{31}=4$ and $q_{32}=6$, then only for $ 3/2 < e < 2$, the rankings are different. 
However, according to Eqn.~\eqref{eqn:compare} as well as Eqn.~\eqref{eqn:toy}, for $ETX=1$ both schemes will put similar weights on the links but with different negative offsets (2 for BCP and 1 for HDCP). Thus, the steady state performance will be same for both but with slightly lesser queue sizes in HDCP, which is also verified by our experiments. 
%
To verify whether the presence of too many perfect links is one of the reason behind the similar performance of HDCP and BCP, we plot the CDF of the ETX traces collected from all the nodes during a real collection experiment, in Figure~\ref{fig:etx_stat}. In Figure~\ref{fig:indiff_curves}, we plot the CDF of the average link costs (average ETX per link) of the shortest paths between every possible pairs of nodes in the testbed. 
Figure~\ref{fig:etx_stat} illustrates that a significant number ($\approx 40\%$) of links are perfect links ($ETX\approx 1$) while Figure~\ref{fig:indiff_curves} implies that approximate $40\%$ of the shortest paths consists of only perfect links ($ETX\approx 1$). Furthermore, approximate $60\%$ of the shortest paths in the network between any possible node pair consists of links with average ETX of $1.25$, as shown in Figure~\ref{fig:indiff_curves}. All these statistics suggest similarity in the rankings of neighbors as well as similarity in performance for the BCP and the HDCP algorithms.
In summary, since we do not observe much of a difference in the performance of the HDCP and the BCP algorithms, we conjecture that in our experiment setup, the probabilities for different rankings of the neighbors (more specifically, top 2 neighbors) are very low. 
 

Next, in order to verify whether similarity in neighbors' ranking will result in similar performance, we perform a theoretical analysis of the steady state queue gradients for both HDCP and BCP. The steady state queue sizes depends on the smallest cost path to the sink. Say, for a node $i$, there exists $k\in \{1, 2, \cdots, K\}$ possible paths and each path consists of one or more links $l_k$. Then the steady state queue size for node $i$ in BCP will be $\overline{w}^{bcp}_i=\min_{k\in \{1, 2, \cdots, K\} } \left[ \sum_{j=1}^{l_k} 2*ETX_{k,j} \right]$ , while in HDCP the steady state queue size will be  $\overline{w}^{hdcp}_i=\min_{k\in \{1, 2, \cdots, K\} } \left[ \sum_{j=1}^{l_k} ETX_{k,j} \right]$, where $ETX_{k,j}$ represents the ETX of the $j^{th}$ link of the $k^{th}$ path from node $i$ to the sink. Thus, we can say that the steady state path to sink for each node in HDCP is same as the BCP. Now, if the packet generation rate is low, every packet will always follow the steady state gradient and, thereby, follow the same path leading to similar performance. Now, if the packet generation rate is high, in worst case we will have a batch arrival of packets at some node, say $i$. Let us assume that when node $i$ disseminate the batch arrival packets, all the neighboring nodes of $i$ are unchanged, i.e., no packet arrival (except from the node $i$)  or departure takes place.
In this situation, node $i$ will keep on transmitting to the neighbor that is part of the best path to sink, until a point when the weight for the respective link becomes worse than the 2nd best link. In the following, we analyze at what point, i.e., after how many packet transmissions, node $i$ will switch to the second best link.
\begin{itemize}
    \item \textbf{BCP:} Node $i$ prefers a neighbor node $m$ over another neighbor node $n$, iff:
    \begin{equation}
    q_i-q_m-2\times etx_{im} > q_i-q_n-2\times etx_{in}
    \end{equation}
    where $q_i,q_m,q_n$ represent the queue sizes at node $i$, $m$, and $n$, respectively and $etx_{im},etx_{in}$ represents the etx of the links $im$ and $in$, respectively.    Now, say after $x$ number of transmissions, the 2nd link is considered:
    \begin{equation}
    \begin{split}
        & \implies (q_i-x)-(q_m+x) - 2\times etx_{im} = (q_i-x)-{q}_n-2\times etx_{in} \\
        & \implies x=q_n-q_m+2\times (etx_{in}-etx_{im})
    \end{split}
    \end{equation}
    Now, WLOG assume that the node $m$ is part of the best path to the sink from node $i$, while node $n$ is part of the second best path. Then, $etx_{im}\approx etx_{in} \implies x\approx q_n-q_m$.
    If $etx_{im} < etx_{in} $, $x=q_n-q_m+2\times (etx_{in}-etx_{im})>q_n-q_m$. On the other hand if $etx_{im}>etx_{in}$ implies  $(q_m <q_n)$ for feasibility of the rankings in focus, which implies $x= q_n-q_m-2\times (etx_{im}-etx_{in}) \leq (q_n-q_m)$. 
    \item \textbf{HDCP:} Node $i$ prefers a neighbor node $m$ over another neighbor node $n$, iff:
    \begin{equation}
    \frac{q_i-q_m}{etx_{im}} > \frac{q_i-q_n}{etx_{in}}
    \end{equation}
    where $q_i,q_m,q_n$ represent the queue sizes at node $i$, $m$, and $n$, respectively, and $etx_{im},etx_{in}$ represent the etx of the links $im$ and $in$, respectively.
    Now, say after $x$ number of transmissions, the 2nd link is considered:
    \begin{equation}
    \begin{split}
        & \implies \frac{(q_i-x)-(q_m+x)}{etx_{im}} = \frac{(q_i-x)-q_n}{etx_{in}} \\
        & \implies x(2-\frac{etx_{im}}{etx_{in}})=q_i \times (1-\frac{etx_{im}}{etx_{in}})-q_m+q_n \times \frac{etx_{im}}{etx_{in}}\\
        & \implies x = q_i\times \frac{(1-\frac{etx_{im}}{etx_{in}})}{(2-\frac{etx_{im}}{etx_{in}})}-q_m \times
        \frac{(1-\frac{etx_{im}}{etx_{in}})}{(2-\frac{etx_{im}}{etx_{in}})} - q_m \times \frac{(\frac{etx_{im}}{etx_{in}})}{(2-\frac{etx_{im}}{etx_{in}})} \\
        &\ \ \ \ \ \ \ \ \ \ \ 
        +q_n \times \frac{\frac{etx_{im}}{etx_{in}}}{(2-\frac{etx_{im}}{etx_{in}})}\\
        & \implies x = \frac{1}{2}\times(1-\mathbf{z})\times (q_i - q_m) + \mathbf{z}\times (q_n-q_m) \mbox{ where } \mathbf{z}=\frac{\frac{etx_{im}}{etx_{in}}}{(2-\frac{etx_{im}}{etx_{in}})}
    \end{split}
    \end{equation}
    Now, WLOG assume that the node $m$ is part of the best path to the sink while node $n$ is the second best path. Similar to BCP, $etx_{im}\approx etx_{in} \implies x\approx q_n-q_m$. It also suggest that in HDCP, the number of transmissions before switching depends on a weighted sum of the queue differential of the best link $(q_i-q_m)$ and the queue differential of the 2nd best and the best neighbor $(q_n-q_m)$, where the weights depend on the ratio $(\frac{etx_{im}}{etx_{in}})$. If $etx_{im}<etx_{in}$, $0.5 \leq \mathbf{z}\leq 1$ that implies more weight on $(q_n-q_m)$ thereby increasing the chances of switching as $q_i\geq \max \{ q_m, q_n \}$ which also implies $x \geq \frac{1+\mathbf{z}}{2}(q_n-q_m) \geq (q_n-q_m) $. On the other hand, $etx_{im}>etx_{in}  \implies  (q_m <q_n)$ for feasibility of the rankings in focus and $\mathbf{z}>1$ that suggests $x= \mathbf{z}\times (q_n-q_m) -\frac{\mathbf{z}-1}{2}\times (q_i - q_m) \leq \frac{1+\mathbf{z}}{2}(q_n-q_m) $. 
\end{itemize}
The above analysis suggests that if the outgoing best and 2nd best link of a node have similar ETX, both BCP and HDCP will switch after exactly same number of transmissions under same queue conditions.
Even in other cases for any particular network, the switching patterns are similar and just switches after slightly different number of transmissions, which is a function of $(q_n-q_m)$. Therefore, the observed performance of BCP and HDCP will be similar.  However, this analysis is pertinent to the fact that both HDCP and BCP have same rankings of the neighbors (atleast best two neighbors) which validates our hypothesis.

Based on the theoretical analysis, we conjecture that the similarity of performance between HDCP and BCP in our testbed experiments is due to similarity of rankings of the neighbors in most of the nodes. 
To verify this conjecture, we perform a Kendall Tau test of the ranking data collected from our real experiment setup, as follows.

\subsection{\textbf{Kendall's Tau Test}}
\label{sec:kendall}
In the previous sections, we observed that the optimized versions of BCP and HDCP with $\beta=1$ are very similar to each other in performance.  We hypothesized that this may be due to similarity in the neighbor rankings for the two protocols. 
In order to verify our hypothesis, we collected a set of routing table snapshots from three representative nodes, located at a one hop, two and three hop distance from the sink, respectively, during a real collection experiment. These snapshots contain the information about their neighbors such as backpressure and ETX information from real experiment. Based on those snapshot values, we calculated the Kendall's Tau distance between the neighbor rankings generated by the weight calculation in BCP on one hand, and the neighbor rankings generated by the weight calculation in HDCP for different values of $\beta$ on the other, for all neighbors that have a positive weight in at least one of the two protocols under comparison. Kendall's Tau distance between two rankings indicates the fraction of pairs that are ordered the same in the two rankings. If it is 0, then the two rankings are identical. Higher values indicate more different rankings. 
\begin{figure}[!ht]
    \centering
     \includegraphics[height=0.4\linewidth, width=0.5\linewidth]{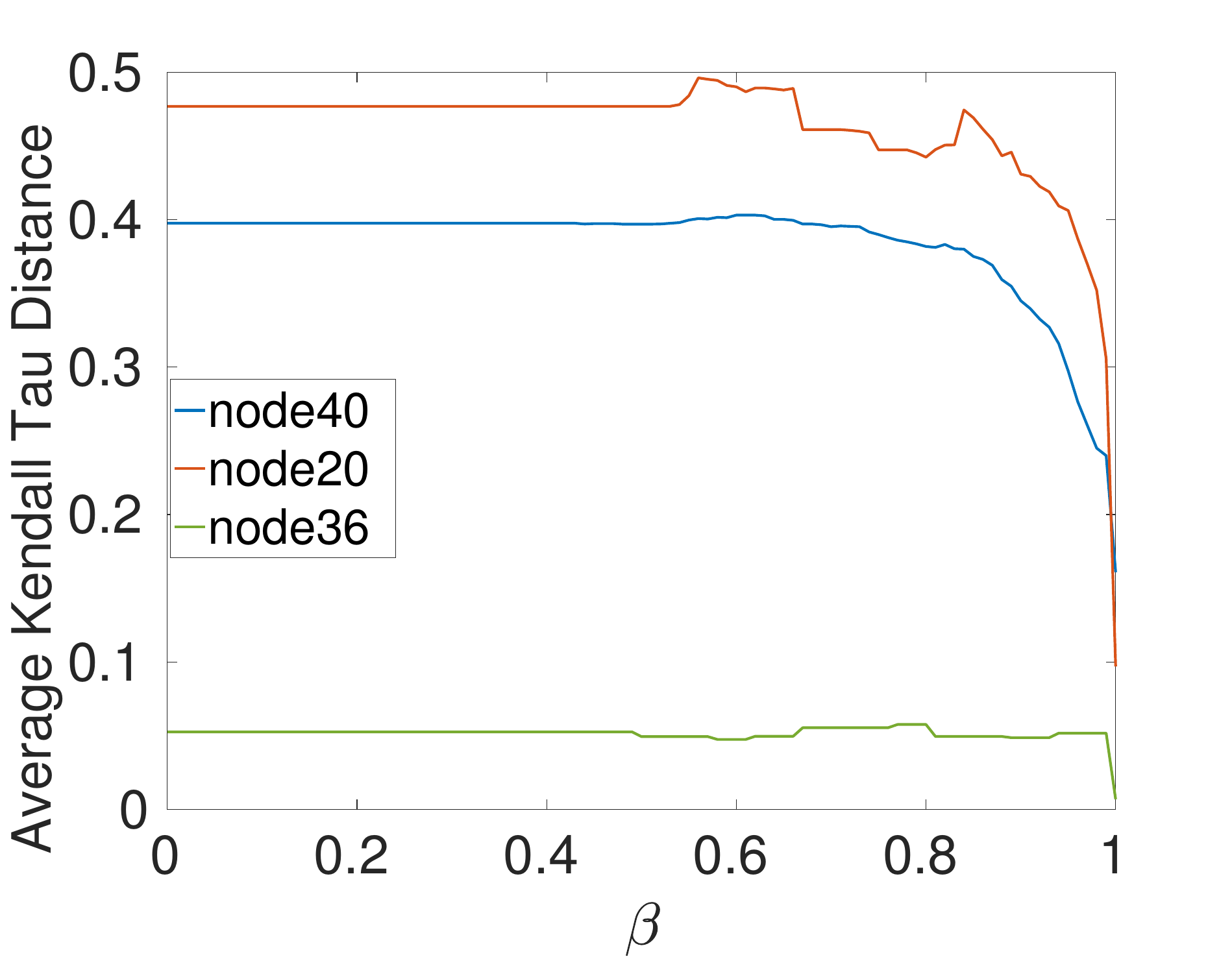}
  \caption{Variation of Kendall's Tau Distance between HDCP and BCP Neighbor Rankings for Different Values of $\beta$}
  \label{fig:kendall}
\end{figure}

We present the results in Figure~\ref{fig:kendall}. It clearly shows that while there is a lack of  correlation for lower values of $\beta$, for $\beta \rightarrow 1$ there is a strong correlation between HDCP and BCP. This verifies our hypothesis and justifies the results shown in this paper.

Another noticeable fact is that for $\beta \in \left[0,0.6\right]$ the Kendall Tau distance with respect to BCP remains almost the same. We performed additional Kendall's Tau correlation analysis between neighbor rankings of HDCP for every possible pair of $\beta \in \{0,0.2,0.4,0.6\}$ and the average distance for each case was found to be less than $0.1$. This is the reason behind the similarity in performance of HDCP with $\beta \in \{0,0.2,0.4,0.6\}$.





\balance
\section{Conclusion}
\label{sec:concl}

We have proposed and implemented a new collection protocol for wireless sensor networks called HDCP \revise{that is the first practical realization of a theoretical algorithm called Heat Diffusion algorithm which is inspired by Thermodynamics}. We have evaluated HDCP on a real wireless sensor network testbed. We have compared the performance of  HDCP with two well-known protocols on this testbed: CTP and BCP.\
Based on the results, we can conclude that HDCP with an optimized parameter setting of $\beta = 1$ performs as well as BCP and outperforms CTP with respect to throughput performance, interference resilience, and low power operation, while all three generally offer about the same end-to-end delay on average in the full throughput region. \

The equivalent performance of HDCP to the previously published BCP is a somewhat surprising finding of this study. From a mathematical perspective, this is not obvious as they employ quite different equations for the weight calculations and indeed in our prior theoretical works Heat Diffusion has been found to perform better than Backpressure scheduling in some respects. But as we have shown, nevertheless, the two protocol implementations provide very similar neighbor rankings in a real network. We believe our finding also lends some support to the notion that it may not be possible to get any higher performance in practice with a dynamic routing protocol that takes into account both queue states and link quality. 


The relative performance of BCP and HDCP in the presence of node mobility is of interest to evaluate in future work, as are extensions of HDCP that can work with IP packets. We would also like to understand how to optimize the various link transmission attempt timers from a more theoretically-informed perspective. 


%


\section*{Acknowledgement}
We would like to thank the US National Science Foundation for their support via award CCF-1423624.



%
{ 
 \bibliographystyle{unsrt}
\bibliography{ref}
}
%




\end{document}